\documentclass[twocolumn,secnumarabic,amssymb,nofootinbib,aps,superscriptaddress,prb]{revtex4-1}

\usepackage{graphicx}
\usepackage{dcolumn}
\usepackage{bm}
\usepackage{xcolor}
\usepackage{amssymb}
\usepackage{wasysym}
\newcommand{\BETS}{$\kappa$-(BETS)$_2$Mn[N(CN)$_2$]$_3$}
\newcommand{\kBETS}{$\kappa$-BETS-Mn}

\newcommand{\eps}{$\varepsilon^{\prime}$}

\begin{document}


\title{Slow and Non-Equilibrium Dynamics due to Electronic Ferroelectricity in a Strongly-Correlated Molecular Conductor}

\author{Tatjana Thomas}
\author{Yassine Agarmani}
\author{Steffi Hartmann}
\affiliation{Institute of Physics, Goethe University Frankfurt, 60438 Frankfurt (M), Germany}

\author{Mark Kartsovnik}
\affiliation{Walther-Meissner-Institut, Bayerische Akademie der Wissenschaften, 85748 Garching, Germany}

\author{Natalia Kushch}
\affiliation{Institute of Problems of Chemical Physics, Russian Academy of Sciences, 142432 Chernogolovka, Russia}

\author{Stephen M. Winter}
\affiliation{Department of Physics and Center for Functional Materials, Wake Forest University, Winston-Salem, North Carolina 27109, USA}

\author{Sebastian Schmid}
\author{Peter Lunkenheimer}
\affiliation{Experimental Physics V, Centre for Electronic Correlations and Magnetism, University of Augsburg, 86135 Augsburg, Germany}

\author{Michael Lang}
\author{Jens M\"uller}
\affiliation{Institute of Physics, Goethe University Frankfurt, 60438 Frankfurt (M), Germany}

\date{\today}

\begin{abstract} 
Using a combination of resistance fluctuation (noise) and dielectric spectroscopy we investigate the nature of relaxor-type electronic ferroelectricity in the organic conductor \BETS, a system representative for a wider class of materials, where strong correlations of electrons on a lattice of dimerized molecules results in an insulating ground state.
The two complementary spectroscopies reveal a distinct low-frequency dynamics. By dielectric spectroscopy we detect an intrinsic relaxation that is typical for relaxor ferroelectrics below the metal-to-insulator transition at $T_{\rm{MI}}\sim 22.5\,$K. 
Resistance noise spectroscopy reveals fluctuating two-level processes above $T_{\rm MI}$ which strongly couple to the applied electric field, a signature of fluctuating polar nanoregions (PNR), i.e.\ clusters of quantum electric dipoles fluctuating collectively. The PNR preform above the metal insulator transition. Upon cooling through $T_{\rm MI}$, a drastic increase of the low-frequency $1/f$-type fluctuations and slowing down of the charge carrier dynamics is accompanied by the onset of strong non-equilibrium dynamics indicating a glassy transition of interacting dipolar clusters, the scaling properties of which are consistent with a droplet model. 
The freezing of nano-scale polar clusters and non-equilibrium dynamics is suggested to be a common feature of organic relaxor-type electronic ferroelectrics and needs to  be considered in theoretical models describing these materials.
\end{abstract}

\maketitle

\section{Introduction}

In contrast to conventional types of ferroelectricity, where the electric polarization is caused by the displacement of ions or the ordering of permanent electric dipoles, the primary order parameter for a novel class of ferroelectrics is determined by electronic degrees of freedom. This so-called electronic ferroelectricity has become the focus of current research interest \cite{Brink2008,Horiuchi2008,Ishihara2010,Kimura2003} and was found, among other systems, in low-dimensional molecular conductors \cite{Lunkenheimer2015a,Tomic2015}. A possible mechanism for the formation of electric dipole moments is charge disproportionation in a modulated or bond-alternated dimerized lattice \cite{Naka2010,Ishihara2014}, which is why the organic dimer-Mott insulators (BEDT-TTF)$_2X$, where BEDT-TTF (in short ET) represents bis(ethylenedithio)tetrathiafulvalene and $X$ an acceptor molecule, are promising candidates for this type of ferroelectricity.
In these systems, the transfer of one electron from two donor molecules ET to a monovalent acceptor molecule $X$ leaves behind a partially-filled molecular orbital, so that the charge carriers responsible for electrial conductivity have $\pi$-hole character with strong electronic correlations due to narrow bandwidth, reduced dimensionality and diminished screening of the charges. 

Prominent examples are the Mott insulators $\kappa$-(ET)$_2$Cu[N(CN)$_2$]Cl (in short $\kappa$-Cl) and $\kappa$-(ET)$_2$Hg(SCN)$_2$Cl, where the $\kappa$-phase denotes a dimerized structure of the ET molecules, exhibiting order-disorder type electronic ferroelectricity that is driven by the charge order within the (ET)$_2$ dimers. 
The possible charge localization on one molecule within the dimer, caused by the interplay of on-site and inter-site Coulomb interactions, leads to quantum electric dipole moments (depicted by red arrows in Fig.~\ref{fig:PNR_cartoon}(b) below). In the first material, ferroelectric order develops from a Mott insulating state and the localization of the charges/spins on the molecules within the dimers was suggested to enable antiferromagnetic (AFM) order making $\kappa$-Cl a multiferroic \cite{Lunkenheimer2012}. The second compound exhibits a metal-insulator transition to a charge-ordered (CO) state \cite{Drichko2014} coinciding with a diverging dielectric constant $\epsilon^\prime$ described by a Curie-Weiss law from which the size of the individual dipoles $p$ due to the intradimer charge disproportionation could be estimated \cite{Gati2018b,Gati2018}. Whereas these ingredients for electronic ferroelectricity can be found in many compounds of organic charge transfer salts, the phenomenology regarding magnetism or electric transport and polarization can be different. For the latter, both long-range order or relaxor-type ferroelectricity, where the peak in $\epsilon^\prime$ becomes suppressed and shifted to higher temperatures with increasing frequency, can arise. The latter behavior, indicative of interacting electric dipoles in the presence of a random potential or competing interactions, has been observed first in the system $\kappa$-(ET)$_2$Cu$_2$(CN)$_3$ \cite{Abdel-Jawad2010}. Likewise, relaxor-type ferroelectricity has been observed in other $\kappa$-(ET)$_2$X compounds, as e.g.\ X = Ag$_2$(CN)$_3$, as well as in systems with other packing motifs or donor molecules, e.g.\ $\beta^\prime$-(ET)$_2$ICl$_2$, $\alpha$-(ET)$_2$l$_3$ and various Pd(dmit)$_2$ systems \cite{Pinteric2016,Iguchi2013,Lunkenheimer2015,Abdel-Jawad2013}. 

In theoretical studies \cite{Naka2010,Hotta2010,Ishihara2014} it has been pointed out that for charge-driven-type ferroelectricity large temporal and spacial dielectric fluctuations are expected which may play a crucial role in the dielectric and optical properties and the ferroelectric phase transition. Besides collective polar charge excitations and the possibility for superconductivity caused by polar charge fluctuations, relevant for this work is the formation of polar nanoregions (PNR) and electronic phase separation, since small polar clusters may cause a relaxor-like ferroelectric dispersion \cite{Ishihara2014}. In the case of the present organic charge-transfer salts (ET)$_2$X, the PNR are considered to consist of short-range domains of electric dipoles localized on the (ET)$_2$X dimers that fluctuate collectively \cite{Abdel-Jawad2010,Hotta2010}. Other important theoretical aspects are the suggestion that charge glassiness is intrinsic to the extended Hubbard model describing the charges on a frustrated triangular lattice \cite{Deglint2022} and that the anomalies in $\epsilon^\prime(\omega,T)$ can be explained by the vicinity of a ferroelectric quantum critical point.\\
In general, it is an important question if relaxor-type ferroelectricity originates from a ferroelectric state, which is broken up into nano-domains due to quenched electric fields, or if it forms from a dipolar glass, i.e.\ freezing PNR \cite{Bokov2006,Fu2009}. Furthermore, the nature of collective excitations in charge-driven ferroelectric phases and their coupling to electric fields, the interplay of dielectric processes associated with domain wall dynamics and the collective response of dipoles, and finally the nature of glassiness are open issues \cite{Tomic2015} 
paramount for an understanding of electronic ferroelectricity in molecular conductors.

In this work, we aim to address these questions by exemplarily studying the relatively new system \BETS\ exhibiting a metal-insulator transition at $T_{\rm MI} \approx 20 - 25$\,K \cite{Kushch2008,Zverev2010}. Besides thermal expansion and resistivity measurements we apply a unique combination of resistance noise spectroscopy and dielectric spectroscopy revealing microscopic evidence for fluctuating PNR as collective excitations and precursors of relaxor ferroelectricity. Their slow dynamics strongly depends on temperature and couples to the electric field. Upon entering the ferroelectric/insulating state we observe a sudden occurrence of strong non-equilibrium dynamics of interacting dipolar clusters or domains consistent with a droplet model of spin glasses. Electric field-dependent and slow (glassy) dynamics therefore is a common theme for a wide class of materials exhibiting electronic ferroelectricity.

\section{Results}
The system chosen for our study is \BETS\ (in short \kBETS), where BETS is the selenium variant of ET in which the four inner sulfur atoms are replaced by selenium. \kBETS\  recently has attracted interest due to the existence of magnetic Mn$^{2+}$ ions in the acceptor molecules, similar to $\lambda$-(BETS)$_2$FeCl$_4$ and $\kappa$-(BETS)$_2$FeBr$_4$ exhibiting magnetic-field-induced superconductivity \cite{Uji2001,Fujiwara2002}. In \kBETS, however, specific heat data \cite{Riedl2021} suggest that the coupling between the BETS spins and the Mn spins is negligible. The weakness of this coupling, which has been also noted in Ref.\ \cite{Vyaselev2017} was estimated based on Shubnikov-de Haas data \cite{Kartsovnik2017} to be smaller than 0.12\,meV, i.e.\ more than an order of magnitude lower than for $\kappa$-(BETS)$_2$FeBr$_4$. 
Based on this finding, theoretical studies including \textit{ab initio} calculations and modeling of magnetic torque \cite{Vyaselev2017} and NMR \cite{Vyaselev2012} results identify a spin-vortex crystal order highlighting the importance of magnetic ring exchange \cite{Riedl2021}. 
The system remains metallic down to $T_\mathrm{MI} \approx 20 - 25\,$K, where it undergoes a metal-to-insulator (MI) transition. The insulating ground state is sensitive to pressure and can be transformed into a superconducting state with $T_\mathrm{c} = 5.7\,$K by applying $p = 0.6\,$kbar \cite{Zverev2010}, reminiscent of the phase diagram of $\kappa$-(ET)$_2$X with polymeric anions \cite{Kanoda1997a}. Band structure calculations \cite{Zverev2010} reveal a narrow bandwidth and a relatively strong dimerization of the BETS molecules, where in the $\kappa$-phase pattern adjacent (BETS)$_2$ dimers are arranged almost orthogonal to each other. 
Quantum oscillations under pressure \cite{Kartsovnik2017,Zverev2019} indicate strong electronic correlations suggesting a Mott instability as the origin of the MI transition. Since in the $\kappa$-phase dimerized systems, the intra-dimer degrees of freedom and charge-lattice coupling play an important role for the electronic ground state properties \cite{Hotta2010,Riedl2022}, the system is another candidate for electronic ferroelectricity. NMR experiments \cite{Vyaselev2011,Vyaselev2012,Vyaselev2012b} have shown that the MI transition is accompanied by a staggered antiferromagnetic order of the BETS molecules' $\pi$-spins, whereas the Mn$^{2+}$ spins, which tend towards antiferromagnetic order, are geometrically frustrated due to the triangular arrangement in the anions. As mentioned above, recent thermodynamic studies in combination with theoretical calculations \cite{Riedl2021} have revealed indications for an unconventional spin vortex state and a decoupling of the $\pi$- and $d$-electrons. This indicates, that the MI transition is not a consequence of magnetic order of Mn$^{2+}$ spins through a $\pi$-$d$ interaction, but a result of strong interactions within the conducting BETS layers itself.\\
In order to study the dynamics related to possible electronic ferroelectricity in single crystals of \BETS, we performed studies of thermal expansion, resistance fluctuation (noise) spectroscopy and dielectric spectroscopy, see the Methods section below.

\subsection{Metal-insulator transition} 
\begin{figure}[t]
	\includegraphics[width=0.47\textwidth]{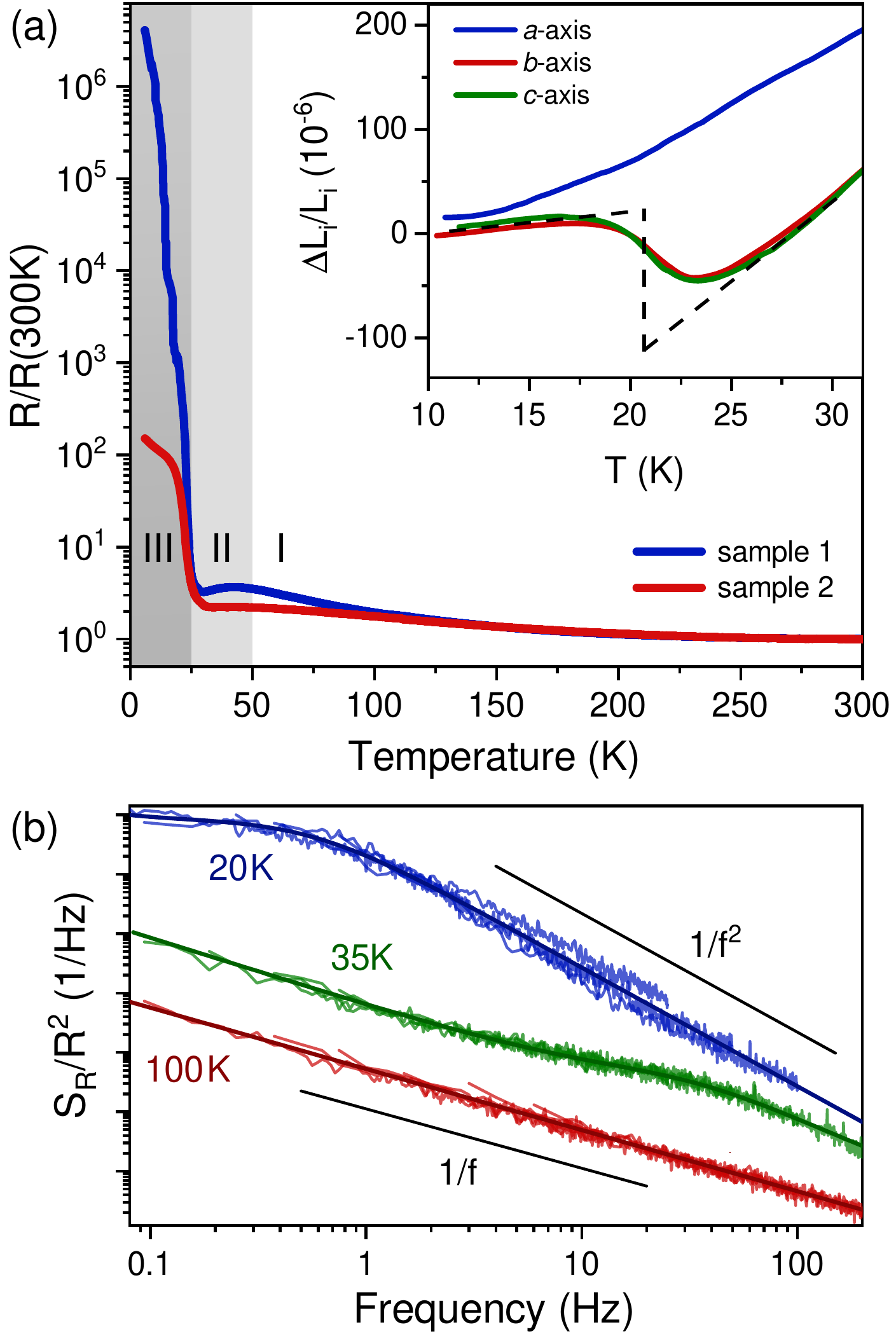}
	\caption{(a) Resistance normalized to the value at 300\,K vs.~temperature of two \kBETS\ samples (\#1 in blue, \#2 in red). The gray shaded regions correspond to the temperature regimes, where pure $1/f$-type spectra (region I), $1/f$-type and superimposed Lorentzian spectra (region II) and spectra of $1/f^2$-type with a strong time dependence (region III) were observed. The inset shows the relative length change $\Delta L_i/L_i$ vs.~$T$ of sample \#3 along different in-plane and out-of-plane crystallographic axes around the MI transition. The dashed line illustrates an idealized sharp jump for the in-plane $b$- and $c$-axis. (b) Typical fluctuation spectra for regions I, II and III in sample \#1. Normalized resistance noise power spectral density, $S_R/R^2(f)$, in a double-logarithmic plot for selected temperatures representing the three temperature regimes. Spectra are shifted for clarity. Different lines shown for each temperature are due to subsequent measurements of different frequency spans. Slopes $S_R/R^2 \propto 1/f$ and $\propto 1/f^2$ are indicated. Solid lines are fits to Eq.~(\ref{eq:Lorentzian}).}
	\label{fig:resistance}
\end{figure}

The resistance vs.\ temperature of two different \kBETS\ samples has been measured perpendicular to the conducting layers upon cooling down the sample with a slow cooling rate of $q \approx -0.7\,$K/min. The temperature evolution of the normalized resistance $R(T)/R(300\,\rm{K})$ is shown in Fig.~\ref{fig:resistance}(a) for samples \#1 (in blue) and \#2 (in red).
At room temperature, the absolute resistance values $R(300\,\rm{K})$ of samples \#1 and \#2 are very similar ($977\,\Omega$ and $970\,\Omega$) with surface areas $A=0.24\,$mm$^2$, $0.17\,$mm$^2$ and crystal thicknesses $d=55\,\mu$m, $45\,\mu$m, respectively, and both samples show a slight increase in resistance upon cooling from room temperature until a small local maximum is observed at about $T\sim42\,$K. A maximum in $R(T)$ was reported previously \cite{Zverev2010} at higher temperatures ($T\sim85\,$K), and has been ascribed to the breakdown of transport coherence perpendicular to the conducting layers and strong electron-phonon interactions \cite{Kushch2008}. The local maximum is also visible in the DC conductivity of our dielectric measurements performed on sample \#4, see Fig.~S9 in the Supplemental Information (SI), at $T \sim 80\,$K in good agreement with \cite{Zverev2010}. Since the sample used for dielectric spectroscopy comes from a different batch, these results emphasize some sample-to-sample dependence of this resistance anomaly.\\
Below a local minimum at $T\sim30\,$K and $34\,$K, respectively, the resistance curves for sample \#1 and \#2 show an abrupt increase by 6 and 2 orders of magnitude, marking the MI transition at $T_{\rm MI} \approx 22.5$\,K defined by a peak in ${\rm d}\ln{R}/{\rm d}T$ \cite{Kushch2008} (with the onset of the resistance increase at 25\,K). This is consistent with the occurrence of signatures of AFM order at $21 - 23$\,K in NMR measurements \cite{Vyaselev2012,Vyaselev2012b}.
The drastically different resistance increase at $T_{\rm MI}$ suggests that sample \#1 is of better crystal quality. 
In the insulating regime, the resistance curve of sample \#1 shows several jumps, which occurred either spontaneously or are due to nonlinear $I$-$V$ characteristics, in particular at the lowest temperatures when reducing the driving current in order to limit the maximum deposited power thereby avoiding Joule heating. 
Below $T \sim 20\,$K, the strong increase of the resistance flattens, which is more pronounced for sample \#2. 
The white and gray shaded areas in Fig.~\ref{fig:resistance}(a) mark three distinct temperature regimes, where different noise characteristics, i.e.\ a distinctly different charge dynamics, was observed (see Fig.~\ref{fig:resistance}(b)), which will be discussed in the next section below.\\
The thermodynamic properties of the MI transition are studied by measuring the thermal expansion of a fourth \kBETS\ sample from a different batch. The results of the relative length changes $\Delta L_i(T) = L_i(T) - L_i(T_0)$, where $T_0$ is a reference temperature (typically the base temperature of the experiment), around the MI transition measured along different in-plane and out-of-plane crystallographic axes $i$ are shown in the inset of Fig.~\ref{fig:resistance}(a). We observe a remarkable anisotropy between the out-of-plane ($a$) and in-plane ($b$ and $c$) axes, where only the in-plane axes exhibit a broadened step-like feature at $T_{\rm{MI}}$, whereas there are no significant changes in the out-of-plane direction. The slightly lower transition temperature for these measurements may be attributed to varying criteria applied for the characteristic temperature in different physical quantities (the onset temperature is roughly the same), or simply may reflect sample-to-sample differences.
The dashed line represents an idealized sharp jump, characteristic for a first-order phase transition. It is worth noting that the jump size of $\Delta L_{b,c}/L_{b,c} \sim 1.3 \times 10^{-4}$ agrees rather well with the volume change calculated by the Clausius-Clapeyron equation using the pressure dependence of $T_{\mathrm{MI}}$ reported in Ref.~\cite{Zverev2010} and the entropy change $\Delta S$ determined in Ref.~\cite{Riedl2021}. 

\subsection{Charge carrier dynamics}
In order to study the charge carrier dynamics around the MI transition at low frequencies, we performed fluctuation spectroscopy measurements in discrete temperature steps using varying setups for different resistance regimes (see ''Methods''). Typical spectra for selected temperatures representing three regimes with distinctly different low-frequency dynamics are shown in Fig.~\ref{fig:resistance}(b). At high temperatures (region I), the resistance noise power spectral density (PSD) shows pure $1/f$-type behavior (red spectrum), i.e.\ $S_R/R^2 \propto 1/f^\alpha$ with the frequency exponent $\alpha = 0.8 - 1.2$. The analysis of these spectra for $T>50\,$K implies enhanced structural dynamics, likely related to the abovementioned peak in the resistivity, and will be published elsewhere. For temperatures below and just above the MI transition, however, a more complex behavior of the charge carrier dynamics emerges. In temperature region II above $T_{\rm MI}$ (light gray shaded area), we observe Lorentzian spectra superimposed on a $1/f$-type 'background' (green), which are caused by dominating two-level fluctuators. 
At temperatures below $T_{\rm MI}$ (region III, dark gray shaded area), we find strongly enhanced Lorentzians with $S_R/R^2 \propto 1/f^2$ above about $f = 1$\,Hz (blue), which are both current dependent and time dependent, see the small offset for subsequent frequency spans for $T = 20$\,K in Fig.~\ref{fig:resistance}(b). This suggests, respectively, (i) a non-linear coupling to the electric field, and (ii) non-equilibrium dynamics and spatial correlations emphasizing the metastable character of the charge carrier dynamics. Therefore, measurements of the so-called 'second spectrum' $S^{(2)}(f_1,f_2)$ (see ''Methods'') have been performed to further analyze the observed ergodicity breaking, see below. We start, however, by discussing the temperature-dependent 'first spectrum' $S^{(1)}(f,T)$, Fig.~\ref{fig:resistance}(b), in regime II (considering at first only the $1/f$-type 'background' without the superimposed Lorentzian contribution) and regime III.\\
\begin{figure}[t]
	\includegraphics[width=0.47\textwidth]{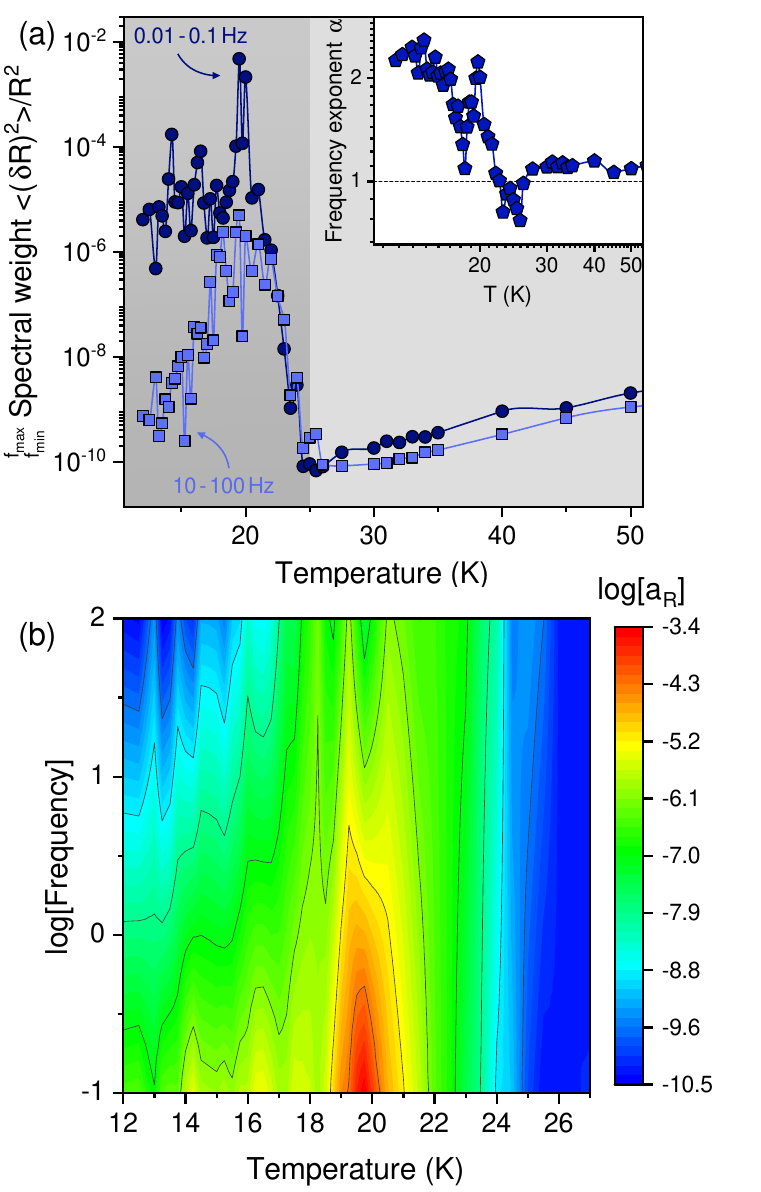}
	\caption{(a) Spectral weight for different frequency ranges ($[0.01,0.1\,$Hz] and $[10,100\,$Hz]) of sample \#1 vs.\ temperature around the MI transition (light grey: regime II and dark grey: III). The inset shows the frequency exponent $\alpha$ vs.~logarithmic temperature. (b) Contour plot of the relative noise level $a_{\rm{R}}= S_R/R^2 \times f$ in log scale in dependence of temperature and logarithmic frequency.}
	\label{fig:noise_lowT}
\end{figure} 
Figure~\ref{fig:noise_lowT}(a) shows the spectral weight $\int_{f_{\rm min}}^{f_{\rm max}}S_R(f)/R^2{\rm d}f$, which corresponds to the variance of the signal for the given bandwidth, vs.\ temperature in a certain frequency interval $[f_{\rm min},f_{\rm max}]$  for two frequency bandwidths $[0.01,0.1\,$Hz] and $[10,100\,$Hz].
Clearly, at all these frequencies the MI transition is accompanied by a drastic increase of the noise PSD's spectral weight by many orders of magnitude, culminating in a maximum at $T\sim20\,$K, which is more pronounced and sharp for the lower frequency bandwidth [$0.01,0.1\,$Hz]. A striking observation is that upon further cooling, after a drastic increase by more than seven orders of magnitude, the noise level at these low frequencies saturates at a high level below the peak (about five orders of magnitude higher than at the onset of the transition), whereas the spectral weight for the higher frequency window [$10,100\,$Hz] decreases upon further cooling to a value comparable to that above the transition. Thus, the charge carrier dynamics below $T_{\rm{MI}}$ is dominated by rather slow fluctuations. The spectral weight of fluctuations with $S_R \propto 1/f^\alpha$ (the noise magnitude) is reflected by the frequency exponent $\alpha(T) = - \partial\ln S_R(T)/\partial\ln f$, shown in the inset of Fig.~\ref{fig:noise_lowT}(a), where $\alpha = 1$ corresponds to a homogeneous distribution of the energies of fluctuators contributing to the $1/f$-type noise, and $\alpha > 1$ and $\alpha < 1$ correspond to slower and faster fluctuations in comparison, respectively \cite{JMueller2018}. 
Upon approaching the MI transition, $\alpha$ being slightly larger than 1 first drops below 1 at about 25\,K (onset of MI transition) representing an initially faster dynamics, before it strongly increases to a peak value $\alpha \sim 2$ at 20\,K, indicating a strong shift of spectral weight to lower frequencies and a drastic slowing down of the dynamics. Below the peak, the fluctuations seem to become faster again upon further cooling but then saturate at consistently high values. A spectrum with $\alpha = 2$ is a signature of non-equilibrium dynamics and often results from a high-frequency tail of a single Lorentzian (cf.\ blue curve in Fig.~\ref{fig:resistance}(b)), which dominates the fluctuations over the entire frequency range of our measurements and implies a switching of the system mainly between two subsequent states. 
The overall picture of an enhanced noise level and drastic slowing down of charge carrier dynamics is highlighted in a contour plot of the so-called relative noise level $a_{\rm{R}}=S_R/R^2 \times f$ (see Fig.~\ref{fig:noise_lowT}(b)) vs.\ temperature and frequency. $a_{\rm{R}}$, a dimensionless quantity characterizing the strength of the fluctuations, upon cooling clearly starts to increase below the onset of the MI transition $T = 25$\,K at all frequencies and peaks at about $T = 19 - 20$\,K. The noise maximum is more pronounced at lower frequencies (note that $a_{\rm{R}}$ and $f$ are shown on a logarithmic scale). In particular, at low temperatures, only the slow fluctuations prevail, whereas at higher frequencies, the noise level of $T > T_{\rm MI}$ is recovered. The noise magnitude for sample \#2 (see Fig.~S1 in the SI) yields qualitatively similar results, whereas the noise increase is not as pronounced and not as sharp as for sample \#1 analogous to the resistance behavior.

\begin{figure}[t]
	\includegraphics[width=0.48\textwidth]{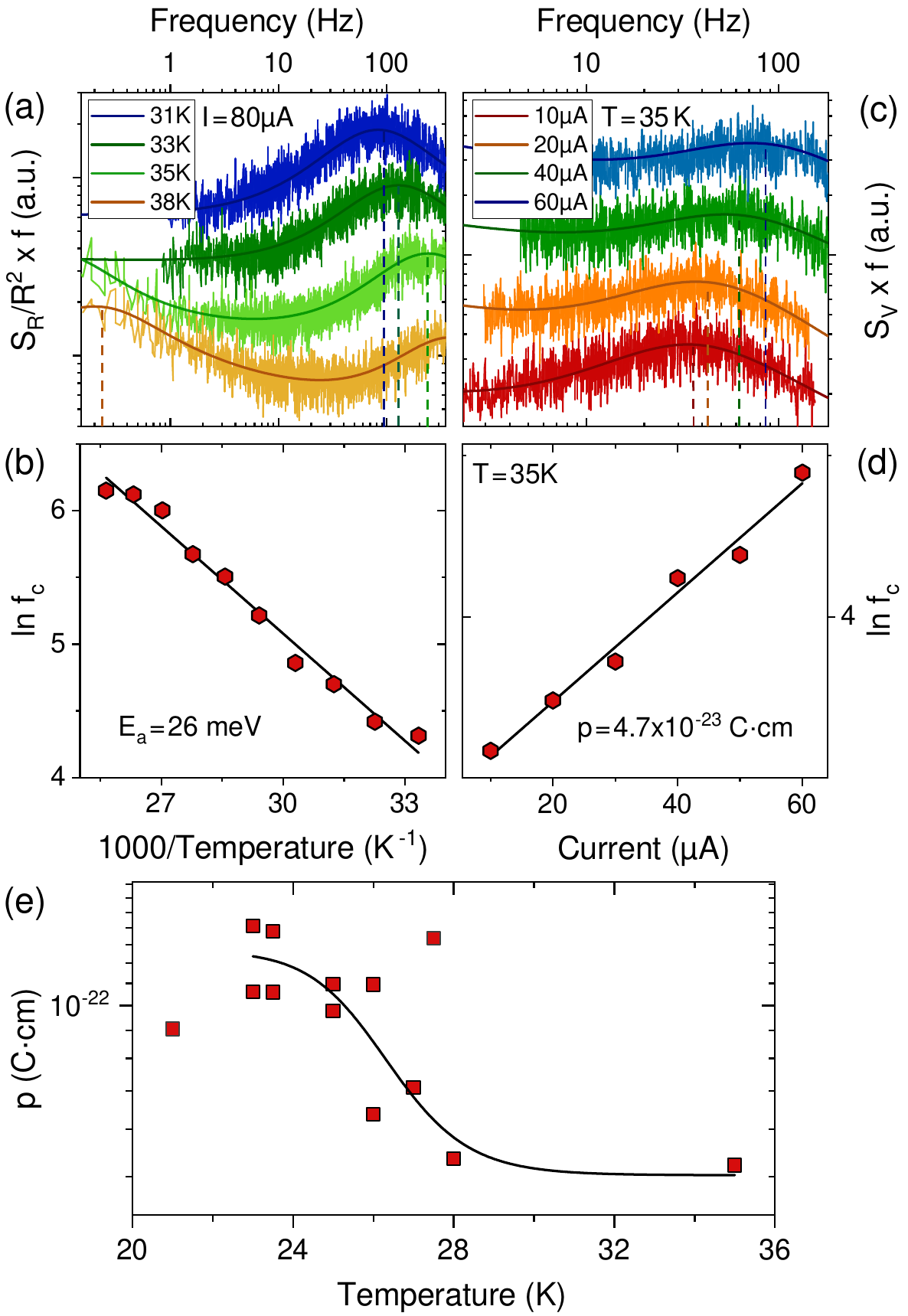}
	\caption{(a) Lorentzian noise contribution as $S_R/R^2 \times f$ vs.\ $f$ superimposed on a $1/f$-type 'background' for different temperatures and constant current ($I=80$\,$\mu$A) for sample \#2. The corresponding shift of the corner frequency given by the maximum of the curves (dashed lines), is shown for the three highest temperatures in (b) vs.\ the inverse temperature with the slope (black line represents an Arrhenius fit) revealing the activation energy. (Dashed line for the 38\,K curve  in (a) indicates the presence of a second Lorentzian peak at lower frequencies, see also Fig.\,S2.) (c) Lorentzian contribution as $S_V \times f$ vs.\ $f$ for $T=35\,$K and different currents, revealing a shift of the corner frequency as illustrated in (d). The dipole moment, which is extracted from the slope of the linear fits according to Eq.\,(\ref{eq:dipole-energy}) shown in (d), is displayed in (e) vs.\ temperature. The black line is a guide to the eye.}
	\label{fig:Lorentzians}
\end{figure}
Now we also consider the Lorentzian spectra superimposed on the $1/f$-type noise and describe the fluctuations by 
\begin{equation}\label{eq:Lorentzian}
\frac{S_R(f,T)}{R^2} = \frac{a}{f^{\alpha}} + \frac{b}{f^2 + f_c^2}
\end{equation}
for temperatures from 50\,K down to 21\,K, i.e.\ the noise peak below the onset of the MIT (regime II), 
exemplarily shown in Fig.~\ref{fig:Lorentzians} for sample \#2,  where $a(T)$ and $b(T)$ are the amplitudes of the $1/f$-type and Lorentzian noise contributions, with frequency exponent $\alpha(T)$ and corner frequency $f_{\rm{c}}(T)$, respectively.
Since observed time dependences of the spectra below $T_{\rm{MI}}$ indicate that the fluctuations are not statistically stationary in regime III,  in the following we discuss the observed systematic non-linear current dependence of the spectra only for temperature regime II. By multiplying the noise PSD with frequency, $S_R/R^2 \times f$, the $1/f$-type contribution becomes a more-or-less constant 'background', thereby emphasizing the Lorentzian contribution with its maximum at the corner frequency $f_{\rm{c}}$. This frequency (marked by dashed lines) shifts as a function of temperature, as shown in Fig.~\ref{fig:Lorentzians}(a) for the corresponding switching process with constant current ($I=80\,\mu$A).  An Arrhenius representation, shown in Fig.~\ref{fig:Lorentzians}(b), reveals thermally-activated behavior $f_{\rm{c}} = f_0 \exp{(-E_{\rm a}/k_{\rm{B}}T)}$ with an attempt frequency $f_0$ of order typical phonon frequencies and an activation energy of $E_{\rm a} = 26\,$meV. We note that the curves at 35\,K and 38\,K in Fig.\,\ref{fig:Lorentzians}(a) reveals the presence of a second Lorentzian peak at lower frequeny. Indeed, in addition to the $1/f$ term, two Lorentzian contributions fit the spectra rather well (dashed lines). Correspondingly, different measurement runs (see SI, Fig.~S2) during warming and cooling reveal various dominating switching processes depending on the temperature: for $T = 35-50\,$K we observe two-tevel switching with $E_{\rm{a}} = 72 - 86\,$meV, whereas for lower temperatures $T=40-25\,$K processes with $E_{\rm{a}}=34-40\,$meV and $E_{\rm{a}}=22-26\,$meV are found, see SI for details. 
As discussed in the following, we assign these two-level processes to thermally-activated switching of clusters of quantum electric dipoles fluctuating collectively, i.e.\ polar nanoregions (PNR).\\
Strikingly, besides the shift with temperature, we observe a strong current dependence of the Lorentzian contribution, see Fig.~\ref{fig:Lorentzians}(c) for another measurement run. At fixed temperature, increasing currents shift the corner frequency to higher values and the magnitude of the Lorentzian $b(T)$ gets suppressed, very similar to previous observations in the square lattice Mott insulator and relaxor ferroelectric $\beta^\prime$-(ET)$_2$ICl$_2$ \cite{JMueller2020}. (Therefore, it is important to note that the temperature-dependent shift discussed above was always analyzed for the maximum applied current.) The linear increase of the corner frequency revealed in a plot of $\ln{f_c}$ vs.\ current, see Fig.~\ref{fig:Lorentzians}(d), indicates that the energy landscape of the thermally activated two-level processes is influenced by the dipole energy $E_{\rm dipole} = p\mathcal{E}$ according to \cite{Raquet2000,JMueller2020}
\begin{equation}
f_{\rm{c}}=f_0\exp\left(\frac{p\mathcal{E}-E_{\rm{a}}}{k_{\rm{B}}T}\right),
\label{eq:dipole-energy}
\end{equation}
where $\mathcal{E}$ denotes the electric field.  The dipole moment $p$ at a fixed temperature can be determined from a linear fit (the slope) 
of the data shown in Fig.~\ref{fig:Lorentzians}(d) yielding $p=4.7\cdot10^{-23}\,{\rm C\,cm}$ at $T=35\,$K, which corresponds to a fluctuating nanoscale polar region of size $100$\,nm (for 3D) or 1000\,nm (for 2D) assuming a charge disproportionation on the dimer of $\delta<0.1e$, similar to $\beta^\prime$-(ET)$_2$ICl$_2$ \cite{Iguchi2013,JMueller2020}. In a 3D scenario the PNR are spherical and in a 2D scenario cylinders with the height of the donor molecule layer are assumed. Assuming 
$\delta<0.005e$ as in $\kappa$-(ET)$_2$Cu[N(CN)$_2$]Cl \cite{Lunkenheimer2012,Lang2014} yields an estimated size of $20$\,nm (3D) or 200\,nm (2D). 
The temperature evolution of $p$ is displayed in Fig.~\ref{fig:Lorentzians}(e) and, despite some scattering in the data,  reveals increasing values for decreasing temperatures, which appear to saturate below  $T_{\rm MI}$. We note that a qualitatively similar behavior is observed for sample \#1 but a clear systematic of the Lorentzian spectra which allowed to track $p(T)$ was only observed for sample \#2.  We speculate that this might be caused by the 
higher degree of disorder for the latter sample,  which may be favorable for the formation of PNR  \cite{Bokov2006}.

\subsection{Dielectric properties}
\begin{figure}[t] 
	\includegraphics[width=0.47\textwidth]{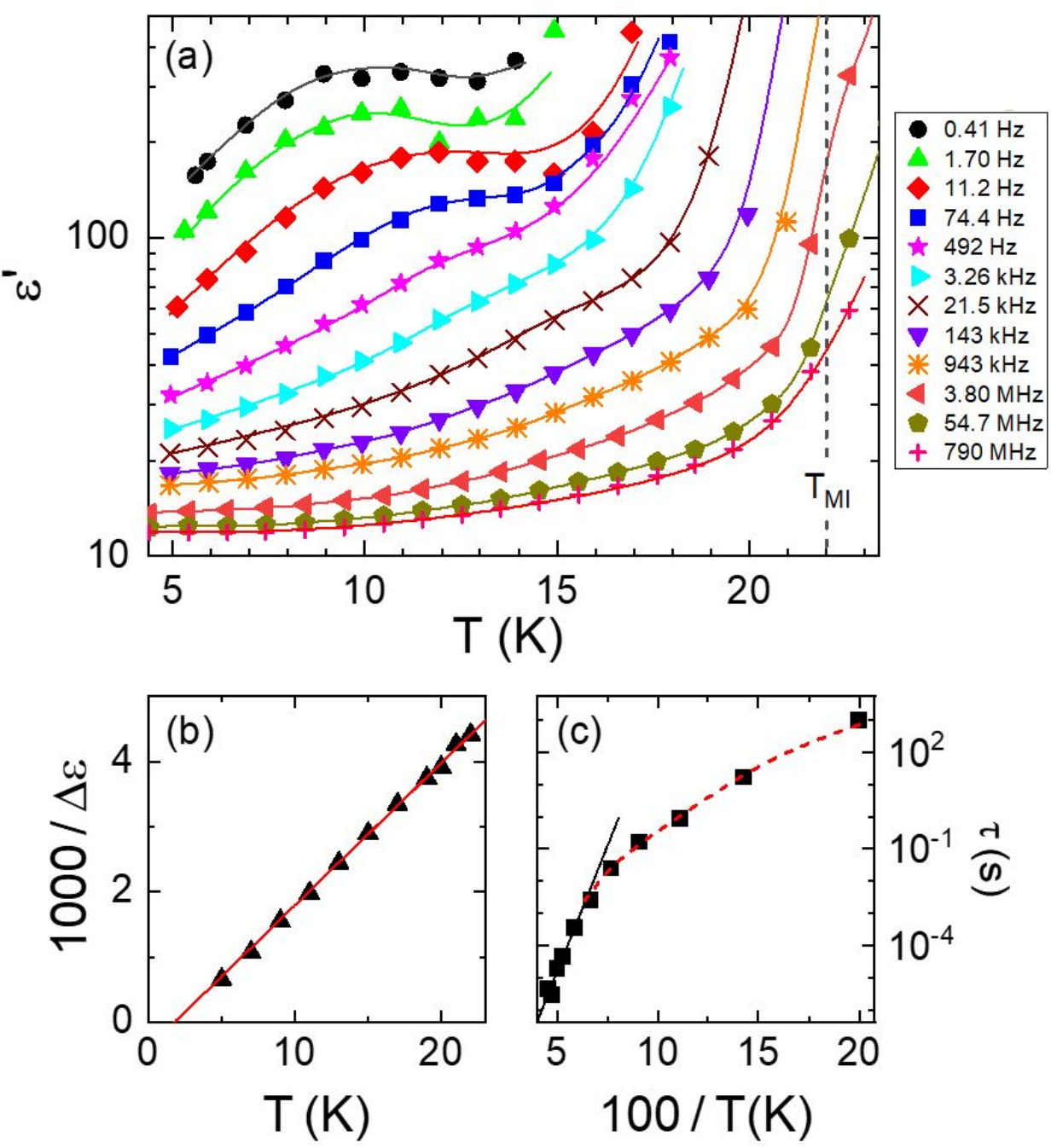}
	\caption{(a) Temperature dependence of the dielectric constant measured at various frequencies. The vertical dashed line indicates the onset temperature of the metal-insulator transition which is at 22\,K for this sample. The lines are guides for the eye. (b) Inverse relaxation strength of the intrinsic relaxation detected at $T<T_{\mathrm{MI}}$ as deduced from fits of the frequency-dependent dielectric data using an equivalent-circuit approach (see SI for details). The line demonstrates Curie-Weiss behavior with $T_{\mathrm{CW}} = 1.8$\,K. (c) Arrhenius representation of the temperature-dependent relaxation time of the intrinsic relaxation process as obtained from the fits. The solid line is a fit of the data above 15\,K by an Arrhenius law with an energy barrier of 31\,meV. The dashed line is a guide to the eye.}
	\label{fig:dielectric_results}
\end{figure}

In order to elucidate the yet unknown electronic state at low temperatures especially with regard to the formation of electronic ferroelectricity, dielectric measurements in a broad frequency range $0.1\, \mathrm{Hz < \nu < 1.8}$\,GHz were performed at temperatures $T = 5 - 300$\,K.  This technique is complementary to the resistance noise spectroscopy discussed above in two ways.  First, dielectric spectroscopy can capture the dynamics of individual electric dipoles, i.e.\ on a microscopic level, fluctuating in an ac electric field and covers a broader frequency range. In contrast, resistance noise spectroscopy captures the dynamics of larger scale polar objects,  PNR or domains,  the fluctuations of which couple to the resistivity.  Second,  whereas dielectric spectroscopy requires a sufficiently insulating behavior of the measured sample, resistance noise spectroscopy works for sufficiently conducting samples. Therefore, complementary information on the dynamics of microscopic objects can be gained from the combination of the two spectroscopies. \\
In the present case,  the relatively high conductivity of \kBETS\ above the MI transition leads to pronounced non-intrinsic effects (so-called Maxwell-Wagner relaxations) \cite{Lunkenheimer2009,Bobnar2002}, probably due to the formation of Schottky diodes at the interfaces between sample and metallic electrodes (see SI for a detailed discussion). Figure~\ref{fig:dielectric_results}(a), showing the temperature dependence of the dielectric constant \eps\ at selected frequencies, thus is restricted to temperatures $T \lesssim T_{\mathrm{MI}}$.
The strong increase of \eps\ at high temperatures is due to the onset of the mentioned Maxwell-Wagner relaxations. At lower temperatures, however, indications for an intrinsic relaxation process are observed, signified by a sigmoidal curve shape, shifting to lower temperatures with decreasing frequency. At the lowest frequencies, a peak develops and \eps\ reaches relatively high values of several hundred. Neglecting the non-intrinsic high-temperature increase, the overall behavior in Fig.~\ref{fig:dielectric_results}(a) is that of the typical dielectric response of relaxor ferroelectrics \cite{Cross1987,Samara2003}.  This makes \BETS\ another example of an organic multiferroic compound. \\
As mentioned above, relaxor ferroelectricity, believed to arise from a cluster-like, short-range ferroelectric order \cite{Viehland1990,Bokov2006,Fu2009}, was also observed in several other charge-transfer salts \cite{Abdel-Jawad2010,Iguchi2013,Lang2014,Lunkenheimer2015,Fischer2021,Canossa2021,Lunkenheimer2015a}. The intrinsic nature of the relaxor-like low-temperature behavior is strongly supported by the similarity of the
results obtained using a different experimental setup, different samples and contact materials as discussed in the SI (Figs.~S5 and S6). To further corroborate the relaxor ferroelectricity in \kBETS, polarization measurements as provided, e.g., in  \cite{Lunkenheimer2015,Fischer2021,Thurn2021} would be desirable. Unfortunately, the rather high conductivity of this material \cite{Kushch2008,Zverev2010} makes such measurements impracticable.\\
Fits of the dielectric spectra, using an equivalent-circuit approach to account for the non-intrinsic contributions, allows deducing the relaxation strength $\Delta \varepsilon$ and the relaxation time $\tau$ of the intrinsic low-temperature relaxation (see SI for a detailed discussion). Figure \ref{fig:dielectric_results}(b) shows the inverse of the resulting $\Delta \varepsilon (T)$. The observed linear increase evidences a Curie-Weiss behavior, $\Delta \varepsilon \propto 1/(T-T_{\mathrm{CW}})$ with $T_\mathrm{CW}=1.8$\,K. $T_\mathrm{CW}$ represents an estimate of the quasistatic dipolar freezing temperature. It is even lower than the already rather low $T_\mathrm{CW}=6$\,K of $\kappa$-(ET)$_2$Cu$_2$(CN)$_3$ \cite{Abdel-Jawad2010} and much lower than in other charge-transfer-salt relaxors where $T_\mathrm{CW}$ values between 35 and 206\,K were reported \cite{Iguchi2013,Lunkenheimer2015,Fischer2021,Canossa2021}.\\
Figure \ref{fig:dielectric_results}(c) shows an Arrhenius plot of the temperature dependence of the relaxation time as derived from the fits. The observed non-linear behavior clearly evidences deviations from simple thermal activation of the detected dipolar dynamics which at best holds for $T>15$\,K only (solid line). In relaxor ferroelectrics, $\tau(T)$ often can be described by the Vogel-Fulcher-Tammann law, which indicates glassy freezing of the dipolar dynamics at low temperatures \cite{Viehland1990}. However, in the Arrhenius representation of Fig.~\ref{fig:dielectric_results}(c), this should lead to an increase of slope with decreasing temperature. Instead, we observe a successively weaker temperature dependence of $\tau$ at low temperatures (dashed line in Fig.~\ref{fig:dielectric_results}(c)). This finding can be explained by a crossover from thermally-activated behavior to quantum-mechanical tunneling below about 15\,K: as the tunneling probability should be essentially temperature independent, a weaker temperature dependence of $\tau$ arises when, upon cooling, thermal activation becomes increasingly unlikely and tunneling phenomena start to dominate.\\
We emphasize again that the relaxational processes observed by resistance fluctuation and dielectric spectroscopy, even though the frequency (and temperature) ranges overlap,  are attributed to different microscopic processes. The former is connected to the switching of PNRs associated with a large effective energy barriers, thus resulting in rather slow dynamics.  The latter captures the reorientation of microscopic dipole moments of the individual (BETS)$_2$ dimers. In combination, the complementary methods provide a more complete picture of the relaxation dynamics in electronic ferroelectrics where disorder and/or competing interactions lead to emergent electronic phase separation. Interestingly, the PNRs form as fluctuating units that can be stabilized in an electric field already above the MI transition and therefore are precursors of the relaxor-type ferroelectricity observed below $T_{\rm MI}$.

\subsection{Nonequilibrium dynamics}
\begin{figure}[t]
	\includegraphics[width=0.47\textwidth]{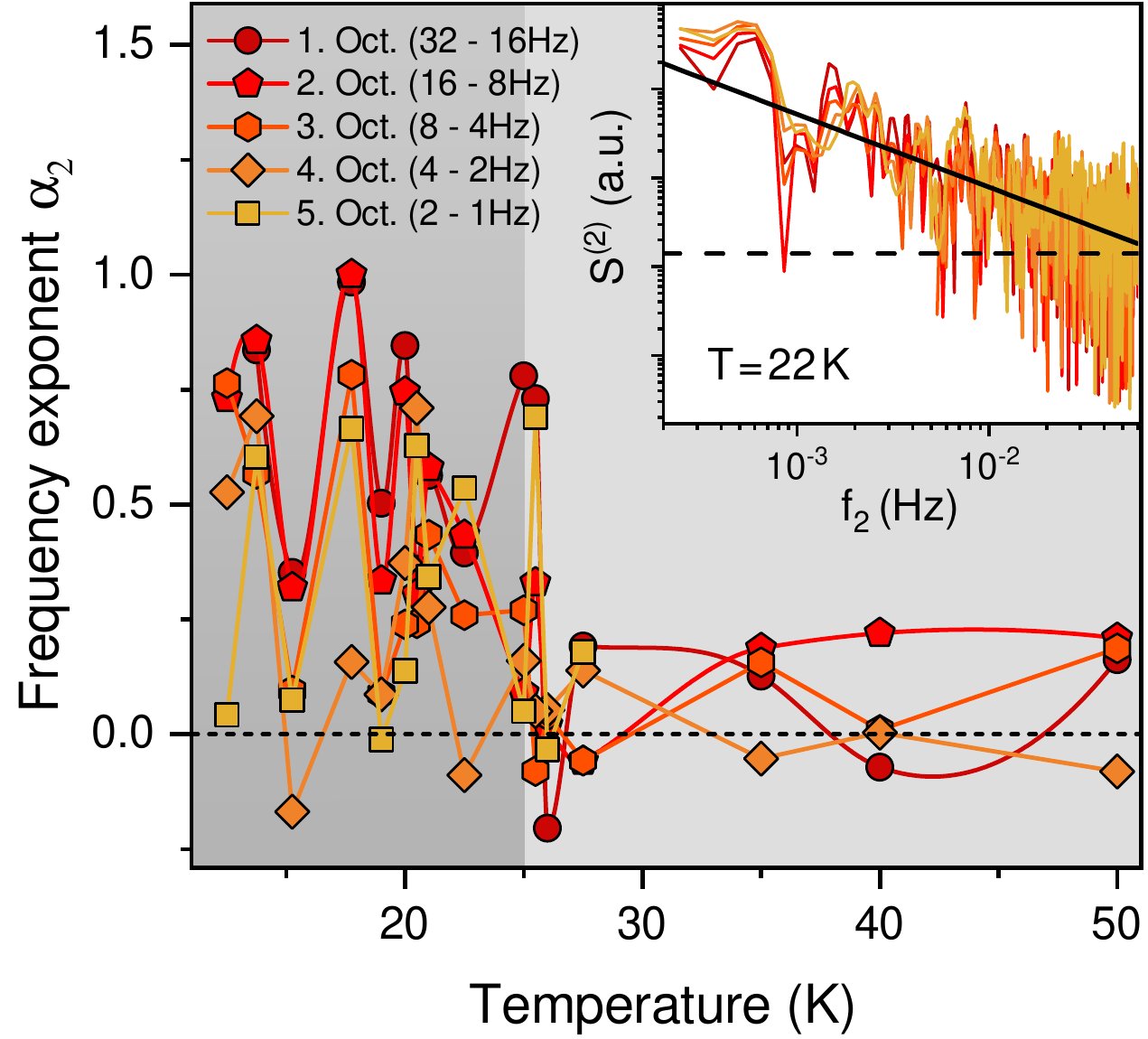}
	\caption{Frequency exponent $\alpha_2$ of the second spectrum $S^{(2)} \propto 1/f_2^{\alpha_2}$ against temperature, which is extracted from a linear fit of the PSD in a double logarithmic plot, as shown in the inset. Different red to orange colors mark different octaves.}
	\label{fig:second_spectrum}
\end{figure}

The results of the dielectric spectroscopy confirm the slowing down of dipolar motion and local ferroelectric correlations for decreasing temperatures, typical for relaxor materials. We now discuss another characteristic feature below the MI transition, i.e.\ a sudden onset of a strong time dependence of the resistance/conductance noise PSD in the insulating state below $T_{\rm MI}$, which results in variations of the spectral weight for repeated measurements at the same temperature, i.e.\ the fluctuations are not  statistically stationary anymore. A similar behavior has been observed for samples \#1 and \#2, whereas for sample \#1, the noise level varies with time by up to one order of magnitude (cf.\ Fig.~S3(a) in the SI), while for sample \#2 the changes are less pronounced. This so-called spectral wandering should be reflected by a non-Gaussian probability distribution of the time signal, which is often caused by spatially correlated fluctuators. 
Indeed, below $T_{\rm{MI}}$, we observe deviations from a Gaussian distribution,  see Fig.~S3(c) in the SI, very similar to the
dynamics of the first-order electronic phase transition in complex transition metal oxides \cite{Ward2009}, which was ascribed to electronic phase separation.  
In order to identify interacting/spatially-correlated fluctuators, we investigated the higher-order correlation function 
by measurements of the second spectrum $S^{(2)}(f_2,f_1,T)$, which corresponds to the PSD of the fluctuating first spectrum (see e.g.\ \cite{Yu2004b} for more detailed information). Here, $f_1 \equiv f$ and $f_2$ correspond to frequencies of the first and second spectrum, respectively, where $f_2$ results from the time dependence of $S^{(2)}(f_1,t)$ at fixed frequency $f_1$. In the case of correlated fluctuators, the second spectrum often shows a frequency dependence according to $S^{(2)} \propto 1/ f_2^{\alpha_2}$ with $\alpha_2 > 0$, whereas $\alpha_2 = 0$ for statistically stationary, Gaussian fluctuations \cite{Weissman1988}.\\
A typical spectrum of $S^{(2)}(f_2)$ is displayed in the inset of Fig.~\ref{fig:second_spectrum} for sample \#1 at $T=22\,$K and indeed reveals a $S^{(2)} \propto 1/f_2$ behavior (black line). The time-dependent spectral weight of the first spectrum, which is used to calculate the second spectrum, is exemplarily shown in the SI [Fig.~S3(b)].
The second spectrum is usually analyzed for different octaves, which are indicated by red to orange colors in Fig.~\ref{fig:second_spectrum} and correspond to varying frequency ranges of the first spectrum (see Fig.~S3(a) in the SI). In order to investigate the development of spatial correlations when crossing the MI transition, we analyzed the frequency exponent $\alpha_2$ of the second spectrum over a wide temperature range, see Fig.~\ref{fig:second_spectrum}. Whereas above the MI transition $S^{(2)}$ is roughly frequency independent ($\alpha_2 \sim 0$), there is a sudden increase of the frequency exponent up to $\alpha_2\sim1$ below $T_{\rm{MI}}$, coinciding with the strong increase in the magnitude of slow fluctuations $S^{(1)}(f)$ shown in  Fig.\,\ref{fig:noise_lowT} above. Clearly, the MI transition is accompanied by strong non-equilibrium charge dynamics indicative of spatially-correlated fluctuators and a dipolar glass.  

\section{Discussion}
From dielectric spectroscopy, the following tentative scenario for the dynamics of the electric dipoles sitting on the (BETS)$_2$ dimers in \kBETS\ is suggested. Upon approaching $T_{\mathrm{MI}}$ from above, due to strong electron correlations the $\pi$-holes become localized on the dimers driving the system into the Mott insulating state going along with a strong increase in both resistance and resistance noise. However, the holes remain delocalized on their respective dimers and the hopping of the holes between the two molecules corresponds to the reorientation of an associated dipolar moment, detected by dielectric spectroscopy.
\begin{figure}[t]
	\includegraphics[width=0.47\textwidth]{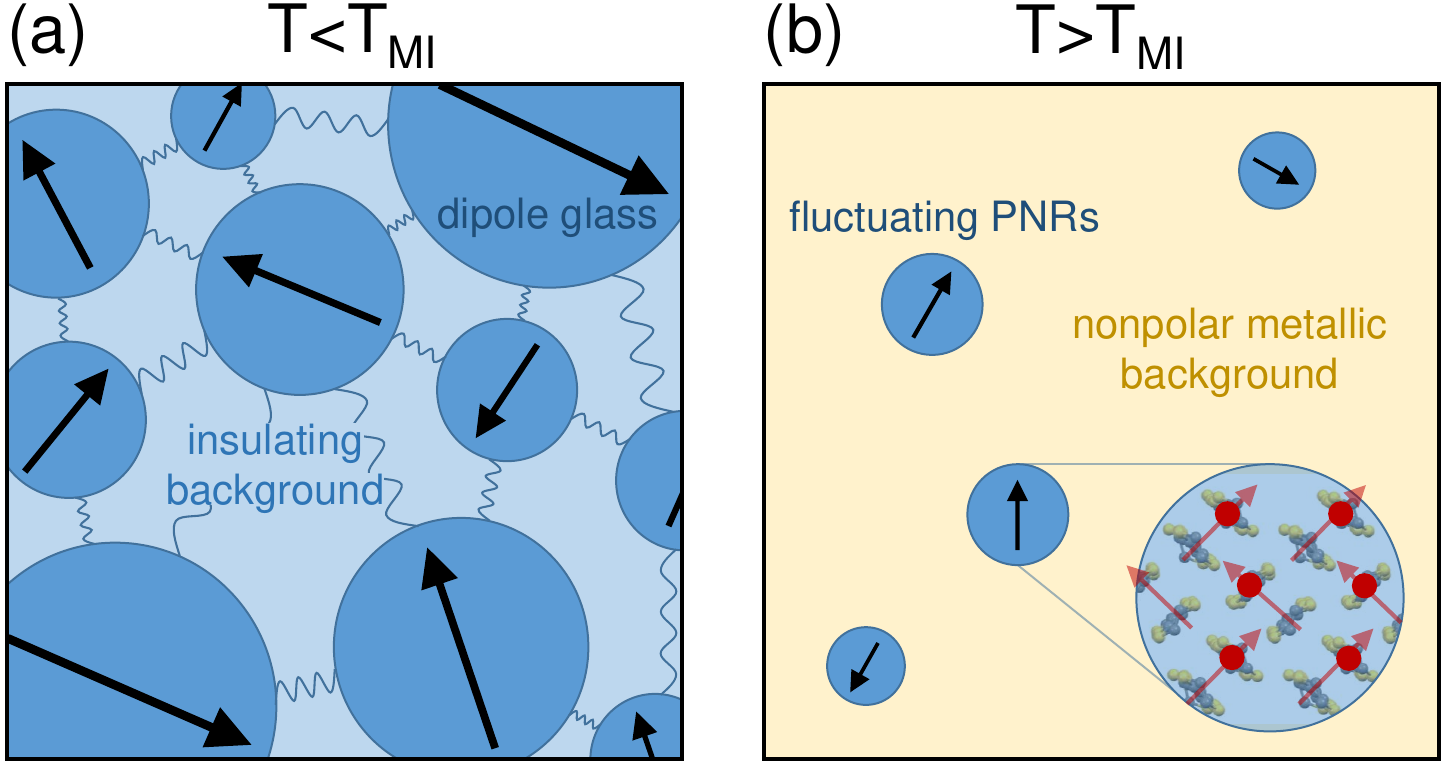}
	\caption{Schematic illustration of the dipolar dynamics that leads to the observed behavior in the resistance noise and dielectric properties: (a) frozen dipole-glass state below $T_{\rm{MI}}$ exhibiting relaxor-type ferroelectricity and strong non-equilibrium dynamics. (b) Polar nanoregions (PNR) preformed in the metallic phase above $T_{\rm{MI}}$. Whereas above $T_{\rm{MI}}$, the PNR are independently fluctuating, they exhibit spatial correlations in the insulating/ferroelectric regime (indicated by wavy lines).}
	\label{fig:PNR_cartoon}
\end{figure}
With decreasing temperature, this local motion starts to slow down and, simultaneously, local ferroelectric correlations --- which, however, partly have been developed already above $T_{\mathrm{MI}}$ --- lead to the cluster-like ferroelectric order, typical for relaxor ferroelectrics. Upon cooling through the MI transition, those PNR which dominate the resistance noise and preexist as fluctuating entities above $T_{\mathrm{MI}}$ undergo a 
transition of interacting two-level systems. 
The abrupt change of the statistical properties of the fluctuations and onset of non-equilibrium dynamics, a signature of spatial correlations, right at the metal-insulator transition can be explained by a change of the background matrix, which is metallic above $T_{\rm{MI}}$ and leads to a screening of the dipole moments. Below $T_{\rm{MI}}$, the insulating background allows the PNR to interact, see Fig.~\ref{fig:PNR_cartoon} for a schematic illustration.\\ 
The strong enhancement of the noise magnitude $S_R/R^2$ at the MI transition at low frequencies below $\sim 500$\,Hz shown in Fig.~\ref{fig:noise_lowT} for sample \#1 can be considered a signature of the first-order phase transition accompanied by emergent electronic phase separation \cite{Chen2007,Ward2009,Daptary2019}. The enhancement of the frequency exponent $\alpha$ corresponding to a drastic shift of spectral weight to low frequencies and slowing down of charge carrier dynamics which persists at low temperatures, is consistent with the localization of charge carriers and the onset of the freezing of PNR switching. The more inhomogeneous transition in the resistivity observed for sample \#2 results in a broadened feature also in the noise magnitude since disorder may lead to the localization of charge carriers in different parts of the sample at slightly different temperatures.\\ 
Moreover, whereas in conventional relaxors the correlated dipolar dynamics arises from ionic motions and exhibits glass-like freezing upon further cooling \cite{Cross1987,Samara2003,Viehland1990}, in \kBETS\ tunneling starts to dominate at low temperatures, preventing the further slowing down and final 
arrest of the essentially electronic dynamics. This may explain the saturation of the noise level/persistence of slow dynamics at low frequencies and low temperatures.
We note that other, more complex scenarios, for example interfacial charge defects within a domain structure arising from coupling of the organic layers to the anion network have been discussed in ref.~\cite{Pinteric2014} in order to explain the relaxor dynamics in $\kappa$-(ET)$_2$Cu$_2$(CN)$_3$. For the present system, however, the existence of PNR and non-equilibrium dynamics is evident.\\ 
Finally, in analogy to spin glasses, the origin of correlated fluctuating PNRs may be classified by the frequency dependence of $S^{(2)}$ vs.\ $f_2/f_1$ in the hierarchical (replica symmetry breaking) or droplet model \cite{Fisher1988a,Fisher1988,Weissman1992,Weissman1993}. The models can be distinguished by their scaling properties, where in a plot $\log{S^{(2)}(f_2,f_1)}$ vs.\ $\log{(f_2/f_1)}$ all curves collaps in the picture of spectral wandering between metastable states related by a kinetic hierarchy. Such a behavior we only observe in close vicinity to $T_{\rm MI}$ (see SI, Fig.~S4). In contrast, for all $T < T_{\rm MI}$ the curves for different octaves are clearly shifted in accordance with a droplet model indicating interacting clusters with a characteristic size distribution, where smaller droplets occur more frequently than large droplets which in turn are more likely to interact. The variation with different octaves can then be understood by long-time measurements of the spectral weight of the first spectrum, where fluctuating processes with time constants of a few hours were observed. This indicates that the noise level switches between a few states only, whose transition times sometimes exceed the measuring time, resulting in different $\alpha_2$ values for repeated measurement runs. Such a behavior was also observed in the dielectric polarization noise of a glassy system \cite{Russell2000} where the spectra continued to change even after several hours due to the large heterogeneity of time constants.\\

In conclusion, our results emphasize the importance of intra-dimer charge degrees of freedom resulting in electronic ferroelectricity in dimerized ET- and BETS-based organic conductors, and demonstrate that the combination of fluctuation and dielectric spectroscopy is a powerful tool to study dipolar dynamics these systems. 
We have identified fluctuating polar nanoregions as precursors of a relaxor ferroelectric state and identified \BETS\ as a new multiferroic material. Upon cooling through the metal-insulator transition non-equilibrium dynamics of interacting PNR emerges consistent with a glassy droplet model. Slow dynamics that depends strongly on temperature and the applied electric field appears as a common theme for a number of different organic dimer Mott insulator compounds where electronic ferroelectricity is discussed, among them $\kappa$-(ET)$_2$Cu[N(CN)$_2$]Cl, $\beta'$-(ET)$_2$ICl$_2$, $\kappa$-(ET)$_2$Hg(SCN)$_2$Cl or $\kappa$-(ET)$_2$Cu$_2$(CN)$_3$, where charge order within the dimers results in an electric polarization. The influence of a random lattice potential, which might cause the glassy/relaxor properties, and of the strength of the coupling of charge (and spin) to the lattice degrees of freedom are aspects worth investigating in the future.

\section{Methods} \label{Methods}

\subsection{Samples investigated}

Single crystals of \kBETS\ with plate-like geometry were grown by electrochemical crystallization \cite{Kushch2008}. In total five different samples (two for noise spectroscopy, one for thermal expansion measurements and two for dielectric spectroscopy) from three different batches were studied. The samples for dielectric measurements (\#4 and \#5) and thermal expansion (\#3) are from the same batch (Garching-SW-03). The samples for noise spectroscopy are from different batches (sample \#1 from batch NKG279 and sample \#2 from AD1).

\subsection{Fluctuation (noise) spectroscopy}

For resistance and noise measurements, $20\,\mu$m-thick Pt wires were attached to the largest surface of the samples, corresponding to the conducting $bc$-plane, by using carbon paste. Two contacts on each side of the sample allowed a four-point measurement perpendicular to the BETS layers. The resistance in the metallic region was measured by an AC technique with a lock-in amplifier (Stanford Research 830), whereas below the MI transition a DC configuration with a Keithley Sourcemeter 2612 was used. Fluctuation spectroscopy measurements were performed with a four-point AC ($T>T_{\rm{MI}}$) and four-point DC method ($T<T_{\rm{MI}}$) by using a preamplifier (Stanford Research 560) and a signal analyzer (Stanford Research 785) (see \cite{JMueller2011,JMueller2018} for more detailed information), which provides the power spectral density of the voltage fluctuations by computing the Fast Fourier Transform. For measurements of the second spectrum, we employed a fast data acquisition card (DAQ) (National Instruments PCI-6281) instead of the signal analyzer in order to determine the power spectral densities from the recorded time signal by a software.

\subsection{Capacitive dilatometry}

Thermal expansion measurements were performed using a homemade capacitive dilatometer with a maximum resolution of $\Delta L/L \geq 10^{-10}$ based on the model of \cite{Pott1983}. The quantity of the relative length change $\Delta L_i(T)/L_i$, with $i=a, b, c$, were determined by $\Delta L_i(T)=L_i(T)-L_i(T_0)$ and the starting temperature $T_0$ of the experiment. 

\subsection{Dielectric spectroscopy}

For the dielectric measurements, gold contacts were thermally evaporated on opposite sides of the plate-like crystals, leading to an electrical-field direction perpendicular to the conducting BETS layers. At frequencies $\nu \lesssim2$\,MHz, the dielectric constant $\varepsilon^{\prime}$, the dielectric loss $\varepsilon^{\prime \prime}$, and the real part of the conductivity $\sigma^{\prime}$ were measured using a frequency-response analyzer (Novocontrol Alpha Analyzer). At higher frequencies, a coaxial reflection technique was used \cite{Boehmer1989} employing a Keysight E4991B impedance analyzer. Temperature-dependent measurements down to $\sim5$\,K were performed using a $^4$He-bath cryostat (Cryovac).

\section*{Data availability}
All the raw and derived data that support the findings of this study are available from the authors upon reasonable request.

\section*{Acknowledgements}
We acknowledge support by the Deutsche Forschungsgemeinschaft (DFG, German Research Foundation) through the Transregional Collaborative Research Center TRR 288 - 422213477 (projects A06 and B02). Work in Augsburg was supported by the DFG through TRR 80. M.K.\ acknowledges support by the DFG through grant no.\ KA 1652/5-1.

\newpage

\begin{center}
\LARGE{Supplementary Information}
\end{center}

\setcounter{figure}{0}

\maketitle

\subsection*{Noise results of sample \#2}\label{noise_sample2}

\begin{figure}[b]
	\renewcommand{\figurename}{FIG. S$\!\!$}
	\includegraphics[width=0.47\textwidth]{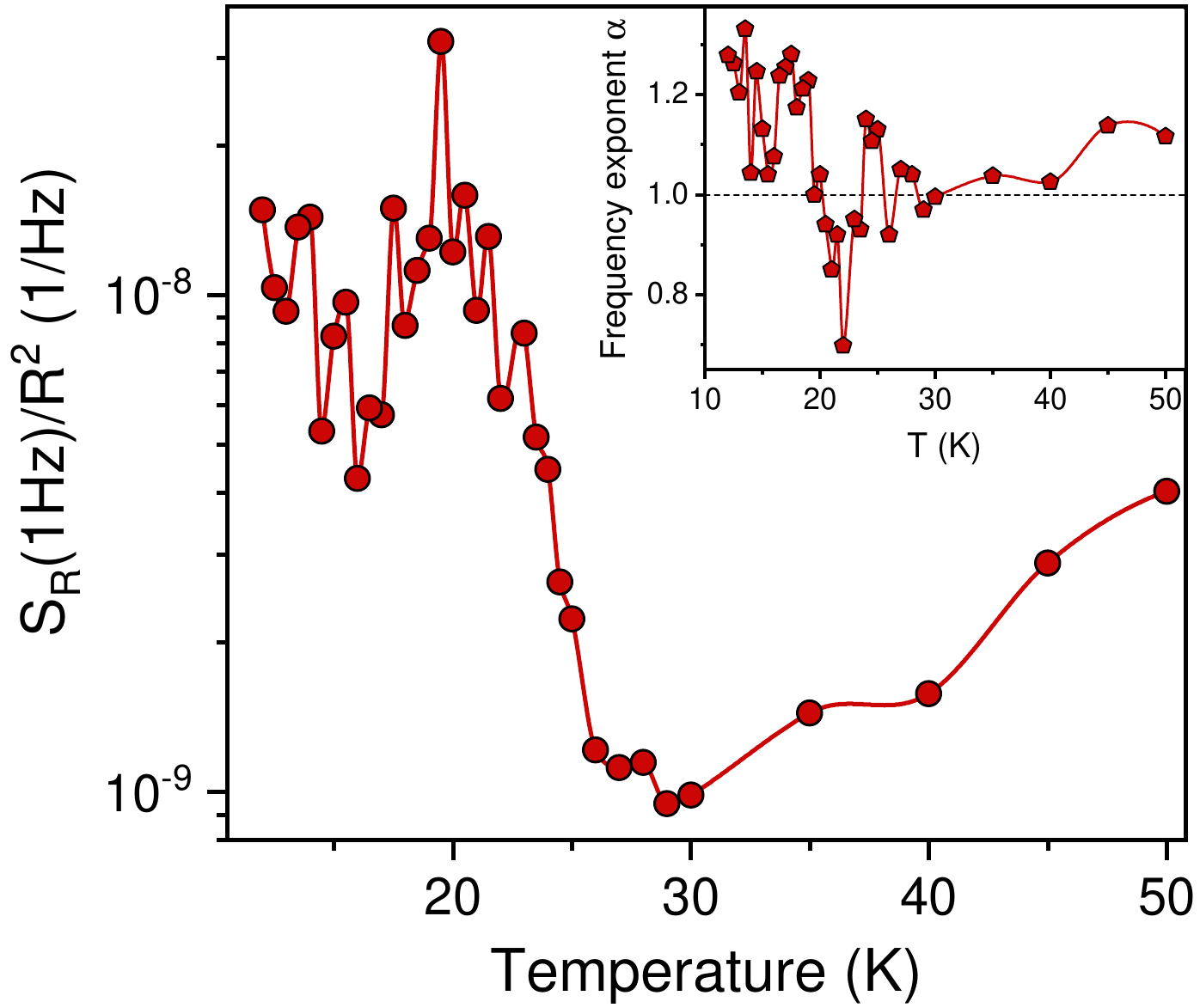}
	\caption{Normalized PSD at 1\,Hz vs.\ temperature around the metal-insulator transition of sample \#2. The inset shows the corresponding frequency exponent $\alpha$.}
	\label{fig:noise_lowT_Zverev}
\end{figure}

\begin{figure}[b]
	\renewcommand{\figurename}{FIG. S$\!\!$}
	\includegraphics[width=0.47\textwidth]{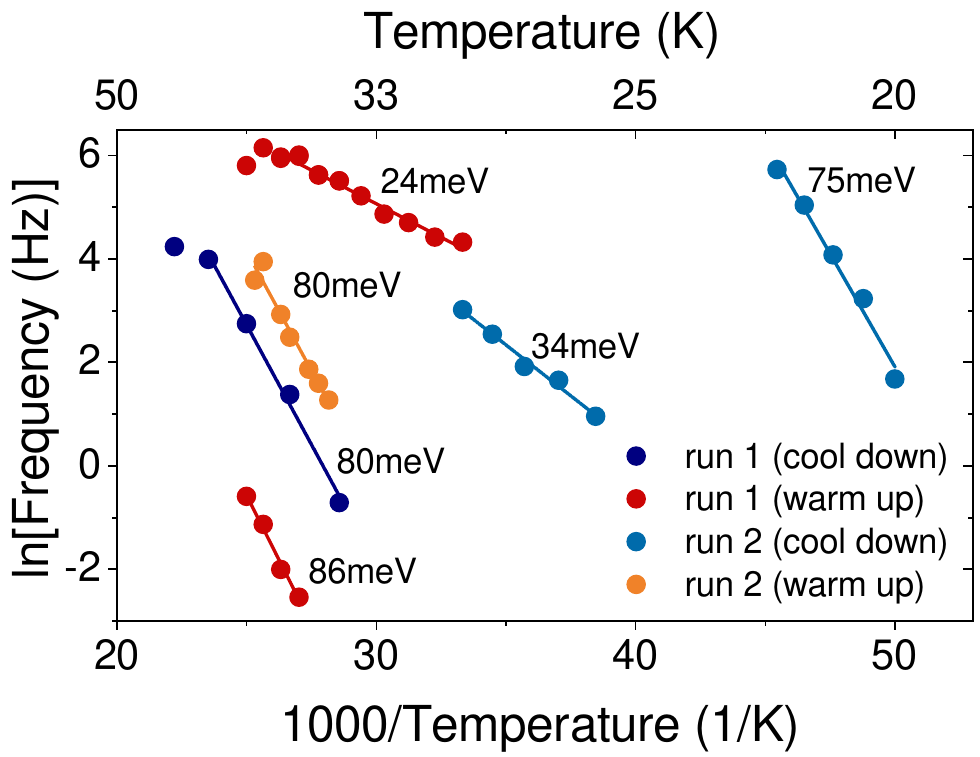}
	\caption{Arrhenius plot of the temperature-dependent corner frequencies of the Lorentzian spectra for different measurements runs in the temperature regime $T=50-25\,$K for sample \#2. Straight lines are linear fits yielding different activation energies of the dominating two-level processes.}
	\label{fig:lorentzian_Tdep}
\end{figure}

We also performed noise measurements on a second sample, which seems to have a poorer crystal quality than the first sample (cf. resistance behavior at the MI transition of both samples, shown in Fig.~1 in the main paper). The normalized resistance PSD at 1\,Hz around and below the metal-insulator transition is shown in Fig.~S\ref{fig:noise_lowT_Zverev} and reveals a strong increase at the metal-insulator transition.
The temperature-dependent noise magnitude is qualitatively similar to the first sample, but neither the increase nor the maximum at $T\sim 20$\,K are as pronounced as for the first sample. The frequency exponent $\alpha$ of the second sample, which is shown in the inset of Fig.~S\ref{fig:noise_lowT_Zverev}, also reveals a similar behavior. At the metal-insulator transition, we observe an increase from $\alpha < 1$ to $\alpha > 1$, which corresponds to a slowing down of the charge carrier dynamics. Compared to the first sample, the frequency exponent in sample \#2 does not reach large values of $\alpha \sim2$. However, the enhancement of $\alpha \sim 1.3$ still demonstrate the existence of dominating slow dynamics in the insulating state. While no pure Lorentzian spectra were observed in sample \#2, many superimposed Lorentzian spectra on a $1/f$ background (cf. green curve in Fig.~1(b)) were found. The analysis of the temperature-dependent Lorentzian contribution for different measurement runs is shown in Fig.~S\ref{fig:lorentzian_Tdep}.
From linear fits of the logarithmic frequency in dependence of inverse temperature, we determine the characteristic energies of the fluctuating processes, which vary from $E_{\text{a}} = 72 - 86\,$meV for $T=35-50\,$K to $E_{\text{a}}=34-40\,$meV and $E_{\text{a}}=22-26\,$meV for $T=40-25\,$K. Therefore, the noise magnitude at 1\,Hz shown in Fig.~S\ref{fig:noise_lowT_Zverev} corresponds to the $1/f$ background without Lorentzian contribution. Furthermore, the spectra of the second sample in temperature regime II show an anomalous current dependence, which still exists in temperature regime III in addition to a time dependence. Since the time dependence is not as pronounced as for sample \#1, it is possible to analyze the systematic current-dependent shift of the corner frequencies of the Lorentzian spectra (see discussion in the main paper). A possible explanation for this and for the less pronounced increase of the noise amplitude as well as the absence of pure Lorentzian spectra for very low temperatures could be a higher degree of disorder in the second sample, leading to an inhomogeneous transition \cite{Chen2007}. Since disorder may favor the formation of PNRs \cite{Macutkevic2011}, this could explain the observation of a large number of superimposed Lorentzian spectra in temperature region II for sample \#2.

\subsection*{Time-dependent first spectrum}\label{Second spectrum}

\begin{figure}[h]
	\renewcommand{\figurename}{FIG. S$\!\!$}
	\includegraphics[width=0.47\textwidth]{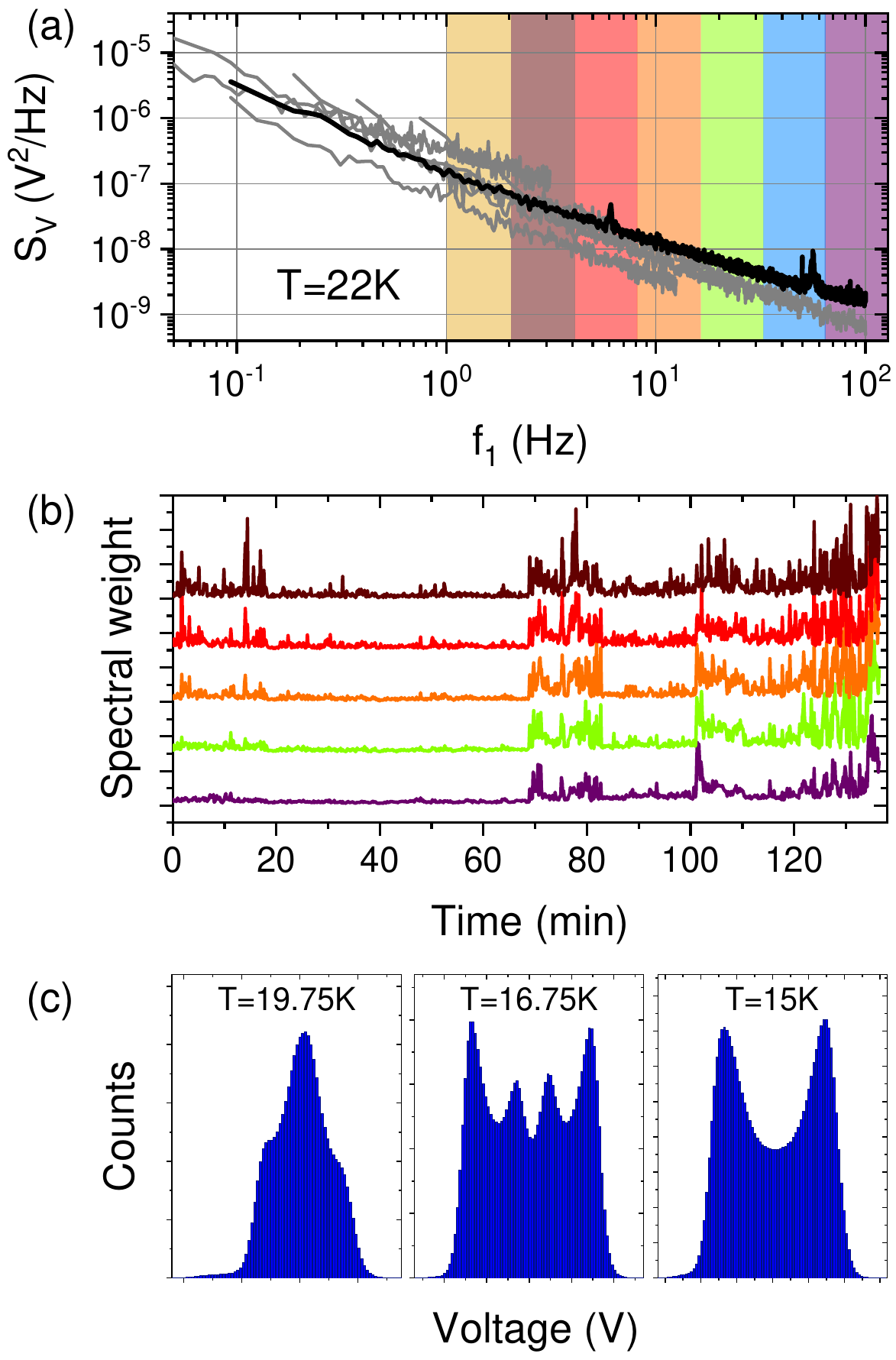}
	\caption{(a) First spectrum measured at $T=22\,$K varying with time as shown by the different frequency spans measured with the signal analyzer (gray) compared to the averaged spectrum measured with the DAQ (black). (b) Noise power of the first spectrum at $T=22\,$K for different octaves in dependence of time. (c) Probabilty distribution of the voltage signal at different temperatures showing deviations from a simple Gaussian function.}
	\label{fig:time_dependence_examples}
\end{figure}

\begin{figure}[h]
	\renewcommand{\figurename}{FIG. S$\!\!$}
	\includegraphics[width=0.47\textwidth]{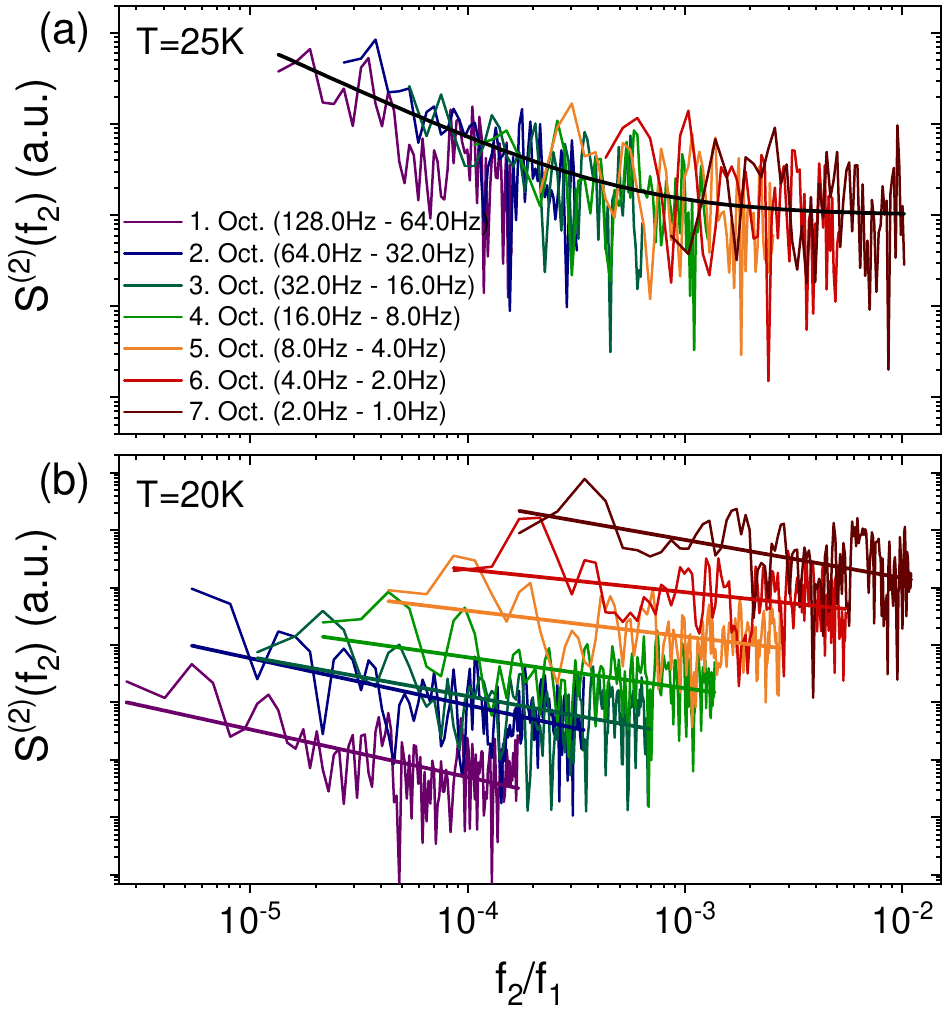}
	\caption{Second spectrum $S^{(2)}(f_2)$ in dependence of $f_2/f_1$ for two selected temperatures. The scaling of the curves for (a) $T=25\,$K indicates correlated fluctuators described by the hierarchical model, whereas the shift of the spectra with decreasing frequencies to higher values, exemplarily shown for (b) $T=20\,$K, implies the description by a droplet model.}
	\label{fig:secondspec_f2overf1}
\end{figure}

The time-dependent first spectrum at $T=22\,$K measured with the signal analyzer (in gray) is shown in Fig.~S\ref{fig:time_dependence_examples}(a) in addition to the averaged first spectrum from the DAQ (in black).
Fig.~S\ref{fig:time_dependence_examples}(b) shows the noise power of the first spectrum as a function of time, where different colors describe different octaves. The measured octaves range from 128\,Hz-64\,Hz (purple) to 2\,Hz-1\,Hz (yellow). A time-dependence of the first spectrum manifests in a non-Gaussian probability distribution of the resistance fluctuations. Examples of the histogram of the voltage signal measured at different temperatures below the MI transition are depicted in Fig.~S\ref{fig:time_dependence_examples}(c) and show clear deviations from a Gaussian distribution. The two maxima for $T=15\,$K indicate a switching mainly between two states, which becomes visible in a frequency exponent $\alpha_1 \approx 2$ at low temperatures (cf. inset of Fig.~2(a)).\\
To classify the origin of correlated fluctuators in the droplet or hierachical model, the second spectrum is analyzed in dependence of $f_2/f_1$, see Fig.~S\ref{fig:secondspec_f2overf1}.
In close vicinity to the metal-insulator transition, exemplarily shown for $T=25\,$K (a), the curves for different octaves collapse, which is in accordance with the hierarchical picture. Contrarily, for temperatures below $T_{\text{MI}}$ (see Fig.~S\ref{fig:secondspec_f2overf1}(b)), we observe a shift for different octaves, whose spectra increase for lower frequencies $f_1$. This behavior is in agreement with the droplet model, where larger droplets occur less frequenctly than smaller ones but are more likely to interact.

\subsection*{Extrinsic and intrinsic dielectric relaxation}\label{relaxation}

\begin{figure}[b]
	\renewcommand{\figurename}{FIG. S$\!\!$}
	\includegraphics[width=0.47\textwidth]{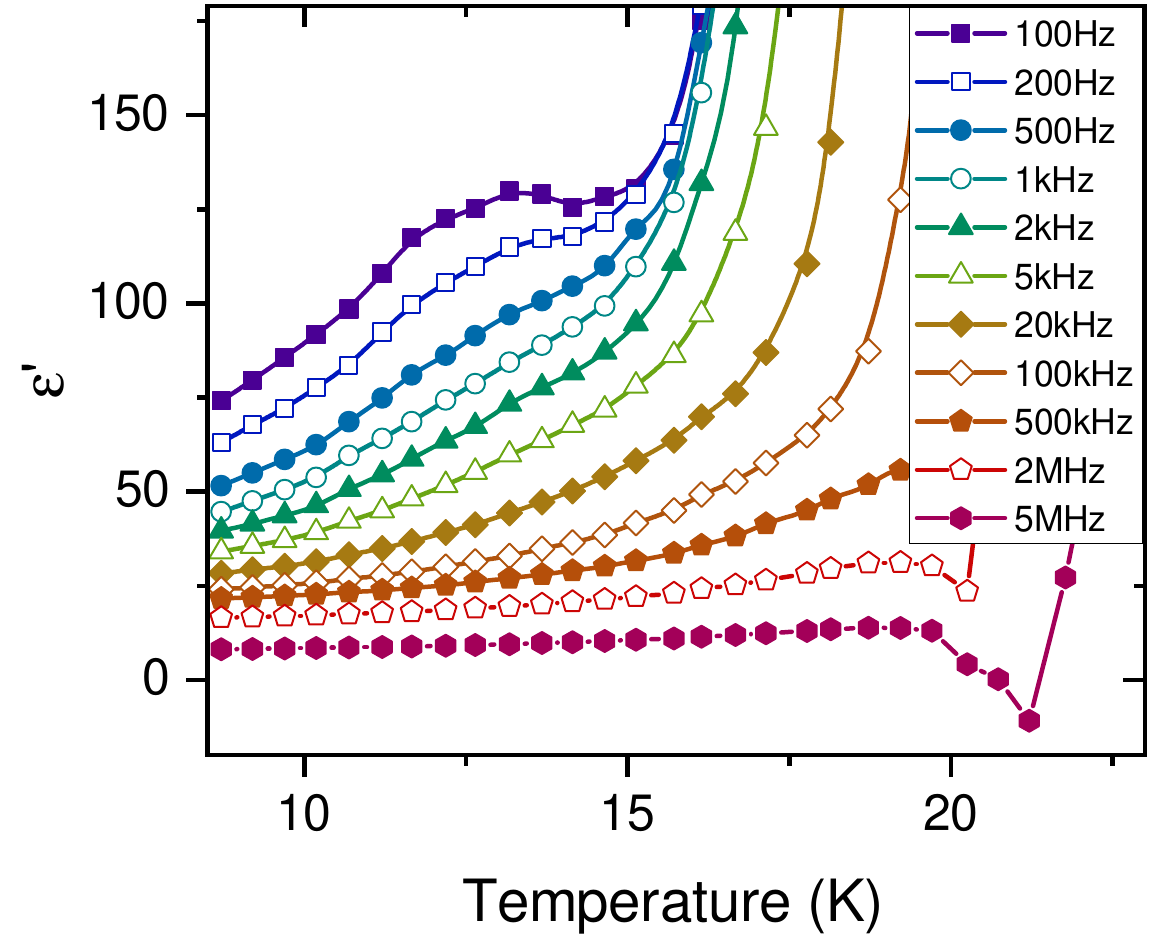}
	\caption{Temperature dependence of the dielectric constant as measured at various frequencies using a different experimental setup than in Fig.~4 but the same sample.}
	\label{fig:dielectric_constant_ffm1}
\end{figure}

\begin{figure}[t]
	\renewcommand{\figurename}{FIG. S$\!\!$}
	\includegraphics[width=0.47\textwidth]{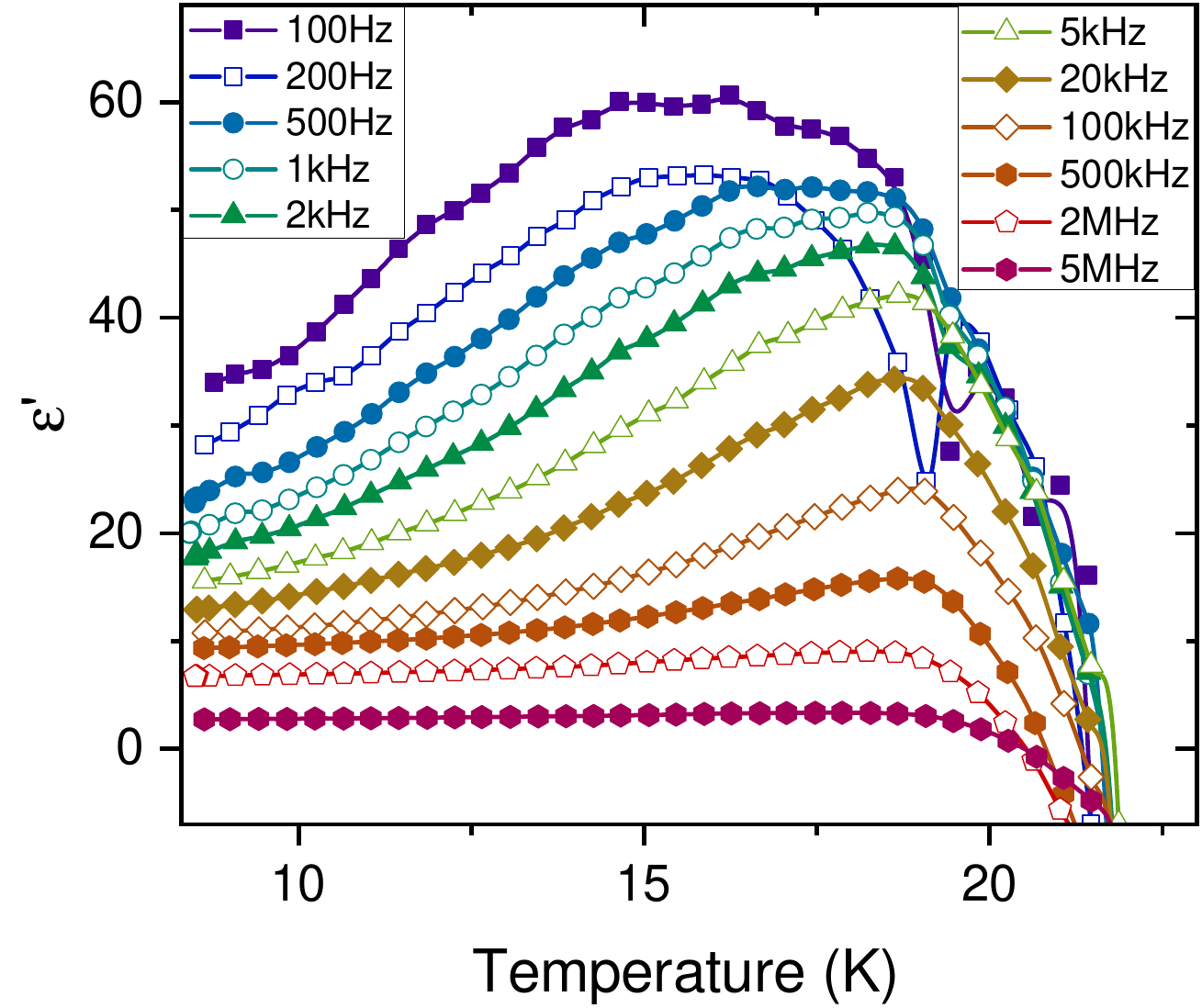}
	\caption{Temperature dependence of the dielectric constant as measured at various frequencies using the same experimental setup as in Fig.~S\ref{fig:dielectric_constant_ffm1} but a different sample with different contact material.}
	\label{fig:dielectric_constant_ffm2}
\end{figure}

\begin{figure}[t]
	\renewcommand{\figurename}{FIG. S$\!\!$}
	\includegraphics[width=0.47\textwidth]{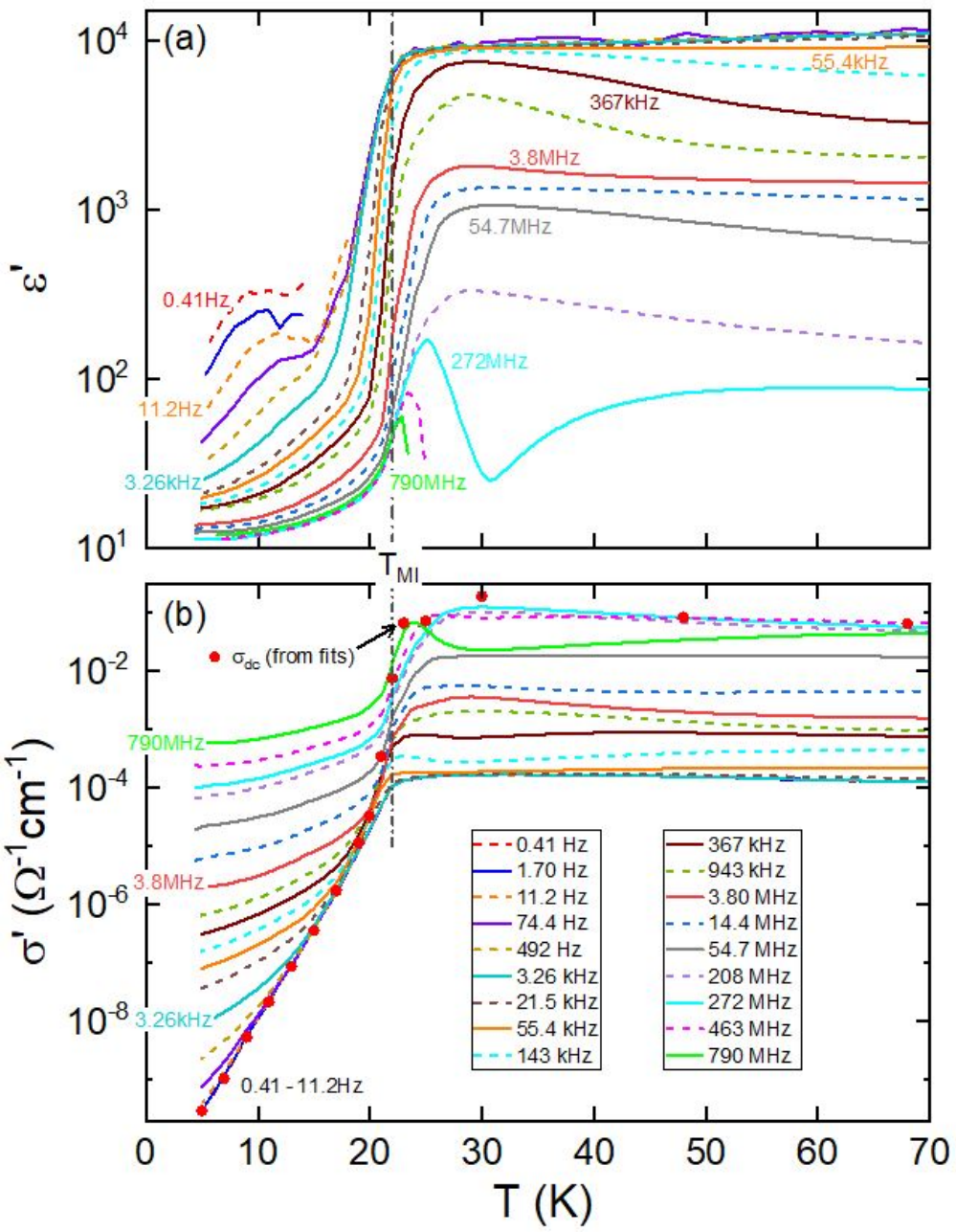}
	\caption{Temperature dependence of the dielectric constant (a) and of the real part of the conductivity (b) shown for selected frequencies. The vertical dash-dotted line indicates the temperature of the MI transition. For the two highest frequencies included in (a), the onset of a steep decline of \eps$(T)$ just above $T_{\mathrm{MI}}$ is due to a crossover to negative values. The filled circles in (b) show the dc conductivity as deduced from fits of the frequency-dependent data using an equivalent-circuit approach \cite{Bobnar2002,Lunkenheimer2009}.}
	\label{fig:dielectric_constant_conductivity}
\end{figure}

Figures S\ref{fig:dielectric_constant_ffm1} and S\ref{fig:dielectric_constant_ffm2} show \eps$(T)$ data at various frequencies and temperatures below and close to $T_{\mathrm{MI}}$, as measured by a different experimental setup than that employed for the results shown in Fig.~4(a) of the main paper. Especially, these measurements were performed using an impedance analyzer (MFIA) from Zurich Instruments covering a frequency range from 1\,mHz to 5\,MHz. Moreover, different samples with different geometries (area-to-surface ratio) and contact materials (carbon paste instead of thermally evaporated gold) were used. The longer unshielded electrical connections that bridge the gap between the end of the coaxial lines within the cryostat and the sample electrodes in the alternative setup, lead to a larger inductance $L$ in series to the sample. As the impedance contribution $\omega L$ (with $\omega = 2 \pi \nu$ the circular frequency) of this inductance increases with frequency, especially at high frequencies RLC resonance effects develop \cite{Lunkenheimer1995,Bobnar2002}.
This explains the non-intrinsic \eps$(T)$ minimum or even negative \eps\ values detected in the vicinity of the metal-insulator (MI) transition in Figs. S\ref{fig:dielectric_constant_ffm1} and S\ref{fig:dielectric_constant_ffm2}. These effects especially occur when approaching the MI transition from below, because there the $\omega L$ contribution can no longer be neglected due to the strong decline of the sample resistance \cite{Kushch2008,Zverev2010}. Irrespective of these effects, both figures exhibit indications of sigmoidal, relaxation-like curve shapes and a peak, resembling relaxor behavior. The similarity to Fig.~4(a) becomes especially evident in Fig.~S\ref{fig:dielectric_constant_ffm1}, which is less affected by the mentioned resonance effects. As noted, e.g., in refs. \cite{Lunkenheimer2002,Lunkenheimer2009}, for non-intrinsic Maxwell-Wagner (MW) relaxations arising from interfacial effects like the formation of Schottky diodes at the electrodes, marked differences in \eps$(T, \nu)$ should arise for measurements of samples with different geometries and contact materials. Here, this is not the case, corroborating the intrinsic nature of the detected relaxor behavior below $T_{\mathrm{MI}}$.\\
Figure S\ref{fig:dielectric_constant_conductivity} shows the temperature dependence of the dielectric constant (a) and conductivity (b) at various frequencies, including data above the MI transition.
These data are from the same measurement run as those shown in Fig.~4(a). It should be noted that $\sigma^{\prime}$ and the dielectric loss $\varepsilon^{\prime \prime}$ are related via $\sigma^{\prime} = \varepsilon^{\prime \prime} \varepsilon_0 \nu$ (where $\varepsilon_0$ is the permittivity of free space), and, thus, Fig.~S\ref{fig:dielectric_constant_conductivity}(b) also provides information on $\varepsilon^{\prime \prime}(T)$. At $T > T_{\mathrm{MI}}$, Fig.~S\ref{fig:dielectric_constant_conductivity} reveals intriguingly complex behavior: peaks show up in both \eps\ and $\sigma^{\prime}$ that shift with frequency and change shape; moreover, at highest frequencies negative values of \eps\ are observed [not directly visible in Fig.~S\ref{fig:dielectric_constant_conductivity}(a) due to the logarithmic \eps\ scale]. Moreover, at low frequencies and high temperatures, unreasonably large values of \eps\ of the order $10^4$ are reached, pointing to non-intrinsic interfacial effects due to electrode-related Schottky diodes and/or internal interfaces \cite{Lunkenheimer2002,Lunkenheimer2009}.\\
Below the MI transition, $\varepsilon^{\prime}(T,\nu)$ [Fig.~S\ref{fig:dielectric_constant_conductivity}(a)] reveals the signatures of relaxor behavior as discussed in the main text. The conductivity $\sigma^{\prime}(T,\nu)$ [Fig.~S\ref{fig:dielectric_constant_conductivity}(b)] here exhibits a relatively smooth decline with decreasing temperature which can be ascribed to the superposition of dc conductivity and the relaxational contribution expected for relaxor ferroelectrics. At $T<T_\mathrm{MI}$, the dc conductivity is approximated by the curve at 0.41\,Hz, from which the higher-frequency curves deviate at different temperatures, depending on frequency. Qualitatively similar behavior was also found for other charge-transfer-salt relaxors \cite{Abdel-Jawad2010,Fischer2021,Lunkenheimer2015}.

\begin{figure}[t]
	\renewcommand{\figurename}{FIG. S$\!\!$}
	\includegraphics[width=0.47\textwidth]{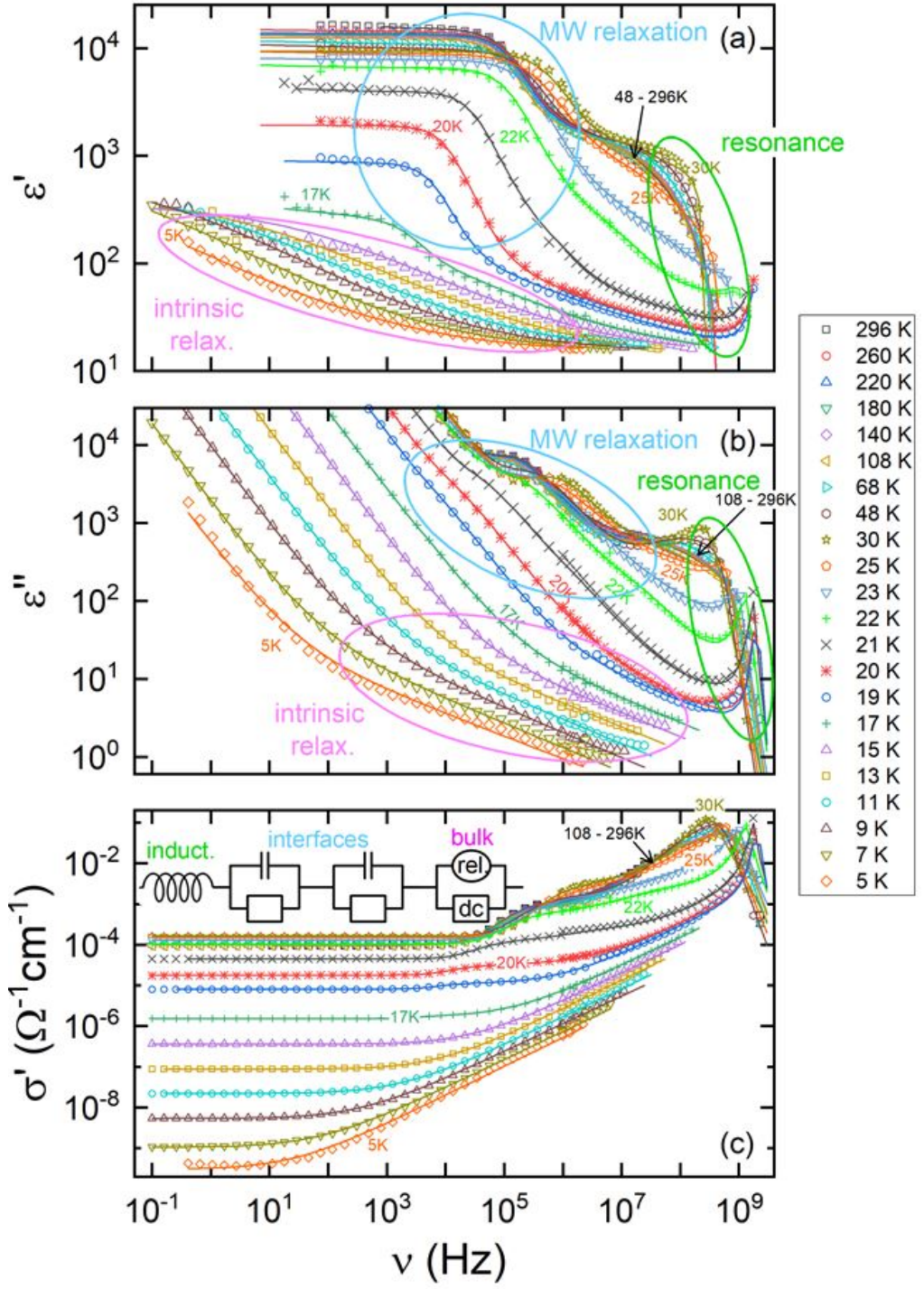}
	\caption{Frequency dependence of the dielectric constant (a), dielectric loss (b), and the real part of the conductivity (c) as measured at selected temperatures. The lines are fits with the equivalent circuit shown in (c) and explained in the text. In (a) and (b), the ellipses approximately indicate the frequency/temperature regions dominated by MW relaxations, resonance effects, and the intrinsic relaxation process at $T<T_{\mathrm{MI}}$.}
	\label{fig:dielectric_equivalent_circuit}
\end{figure}

To quantitatively analyze dielectric data that are assumed to be partially governed by non-intrinsic contributions, fits of the dielectric spectra using an equivalent-circuit approach have to be performed \cite{Lunkenheimer2002,Lunkenheimer2009}. Figure S\ref{fig:dielectric_equivalent_circuit} shows spectra of \eps, $\varepsilon^{\prime \prime}$  and $\sigma^{\prime}$ at various temperatures, obtained from the same measurement run as the data in Figs. 4 and S\ref{fig:dielectric_constant_conductivity}. Again, complex behavior shows up, especially at temperatures above and close to the MI transition. For example, at high frequencies and temperatures negative values of \eps\ and relatively sharp peaks in $\varepsilon^{\prime \prime}$ and $\sigma^{\prime}$ appear, pointing to resonance effects due to the inductance of the rather well-conducting sample [green ellipses in Figs. S\ref{fig:dielectric_equivalent_circuit}(a) and (b)] \cite{Bobnar2002}. The lines in Fig.~S\ref{fig:dielectric_equivalent_circuit} are fits with the equivalent circuit indicated in Fig.~S\ref{fig:dielectric_equivalent_circuit}(c). It includes a series inductance and two distributed parallel RC circuits \cite{Lunkenheimer1996,Lunkenheimer2009,Emmert2011} to account for non-intrinsic interfacial effects. The bulk dielectric response comprises the intrinsic dc conductivity of the sample and an intrinsic relaxation process. The latter is modeled by the empirical Cole-Cole function \cite{Cole1941}  and was only employed in the fits of spectra at $T=22$\,K. At higher temperatures, all the observed complex behavior can be well fitted by the equivalent circuit without assuming any unusual temperature dependence of the involved elements and without intrinsic relaxation process. The steps in \eps$(\nu)$, revealing huge absolute values at their low-frequency plateaus, and the corresponding loss peaks [blue encircled regions in Figs. S\ref{fig:dielectric_equivalent_circuit}(a) and (b)] are due to the fact that the low-conducting regions (e.g., depletion layers of Schottky diodes) become successively shorted by the corresponding capacitances at high frequencies \cite{Lunkenheimer2002,Lunkenheimer2009}. In a transition region between 17 and 22\,K, MW and intrinsic relaxations overlap and had to be simultaneously taken into account by the fits. For $T\leq15$\,K, the bulk conductivity is sufficiently low making electrode effects negligible. Thus, there the intrinsic relaxation [magenta ellipses in Figs. S\ref{fig:dielectric_equivalent_circuit}(a) and (b)] and dc conductivity alone were sufficient to describe the experimental data.\\
It should be noted that the fits of the higher-temperature data involve quite many parameters, which partly are correlated. The aim of these fits is to demonstrate that the very complex temperature and frequency dependence in this region in principle can be described by rather trivial non-intrinsic effects and is not related to any spectacular intrinsic dielectric properties of the sample. For example, within this approach the apparently unsystematic variation of the peaks arising in $\sigma^{\prime}(T)$ [Fig.~S\ref{fig:dielectric_constant_conductivity}(b)] is caused by the temperature variation of the bulk-conductivity derivative switching from $\partial \sigma_{\mathrm{dc}} / \partial T>0$ (below $T_\mathrm{MI}$) to $\partial \sigma_{\mathrm{dc}} / \partial T<0$ (up to about 80 K) and back again to $\partial \sigma_{\mathrm{dc}} / \partial T>0$ (beyond 80 K) \cite{Kushch2008,Zverev2010}. This directly affects the MW relaxation time $\tau_{\mathrm{MW}}$ which is proportional to $1/ \sigma_{\mathrm{dc}}$ \cite{Lunkenheimer2009}, and the resulting back-and-forth shift of $\tau_{\mathrm{MW}}(T)$ leads to the apparently intriguing dielectric behavior at $T<T_{\mathrm{MI}}$.
\begin{figure}[t]
	\renewcommand{\figurename}{FIG. S$\!\!$}
	\includegraphics[width=0.47\textwidth]{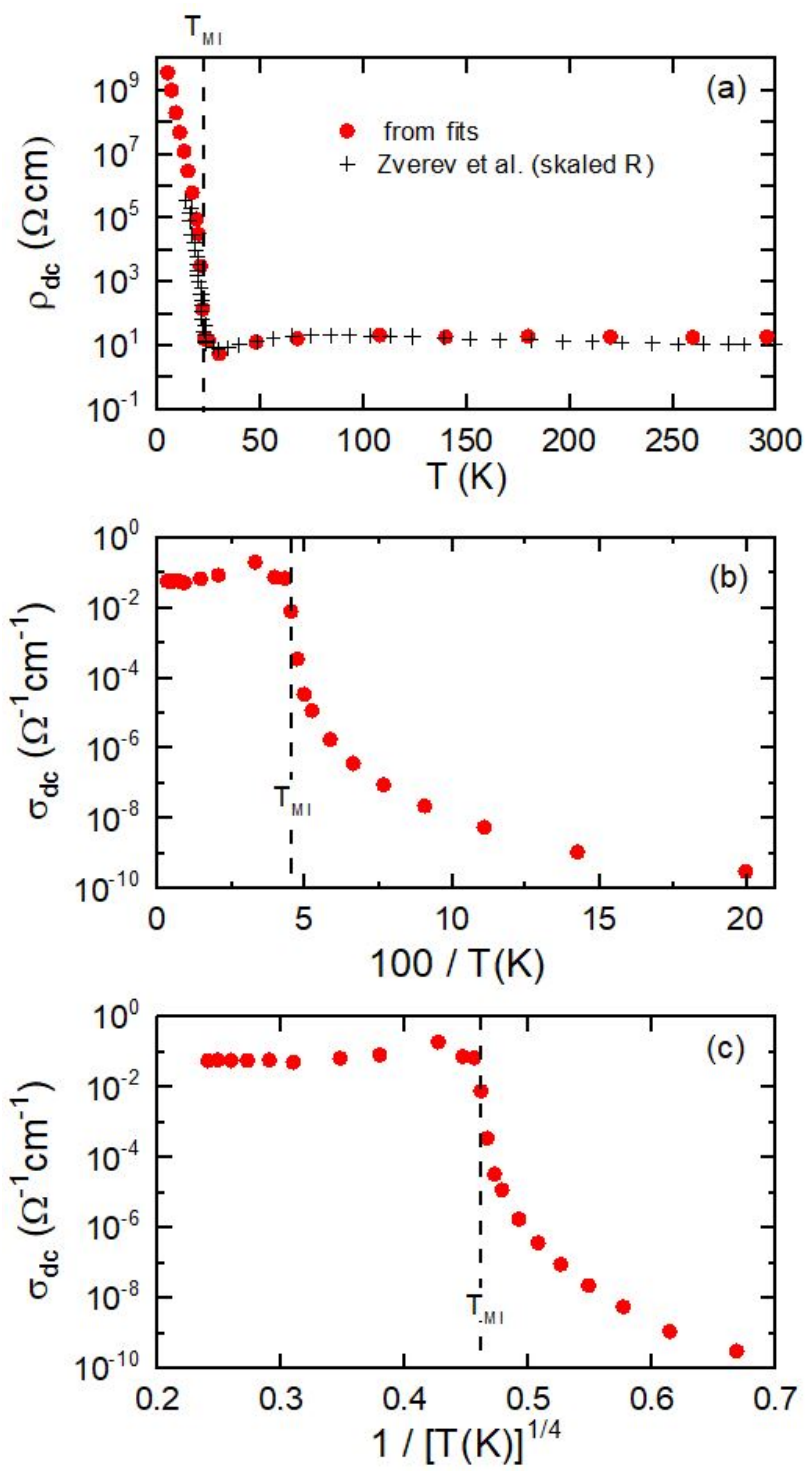}
	\caption{(a) Temperature dependence of the intrinsic dc resistivity as obtained from the fits shown in Fig.~S\ref{fig:dielectric_equivalent_circuit} (circles). The pluses show the results from dc measurements by Zverev \textit{et al}. \cite{Zverev2010}. As there no absolute values were provided, these data were scaled, achieving a reasonable match with the fit results. (b) Present dc-conductivity data shown in an Arrhenius plot. (c) Same data shown in a representation that should linearize $\sigma_{\mathrm{dc}}(T)$  for variable range hopping.}
	\label{fig:resistance_lunkenheimer}
\end{figure}
\newline In addition to the relaxation time and amplitude of the intrinsic relaxation, presented in Figs. 4(b) and (c), another important property derived from the fits is the intrinsic dc conductivity. It is shown for $T<70$\,K in Fig.~S\ref{fig:dielectric_constant_conductivity}(b) by the red closed circles. As mentioned above, for $T<T_\mathrm{MI}$, it agrees with $\sigma^{\prime}(T)$ at the lowest frequency. Above the MI transition, the measured $\sigma^{\prime}$ is lower than $\sigma_{\mathrm{dc}}$ which reflects its non-intrinsic nature in this region, due to interfacial effects. The dc resistivity, $\rho_{\mathrm{dc}} = 1/\sigma_{\mathrm{dc}}$, is shown up to room temperature in Fig.~S\ref{fig:resistance_lunkenheimer}(a) (circles). It is well consistent with $R_{\mathrm{dc}(T)}$, measured by Zverev et al. using a four-probe technique (plus signs) \cite{Zverev2010}. This nicely supports the validity of the performed equivalent-circuit analysis. Interestingly, due to the superior resolution of dielectric devices at high resistances, the present dc results extend to considerably higher resistivities and lower temperatures than the reported four-point measurements. This enables a meaningful check of the temperature dependence of the dc resistivity (or conductivity) below $T_{\mathrm{MI}}$ for thermally activated behavior. Figure S\ref{fig:resistance_lunkenheimer}(b) shows $\sigma_{\mathrm{dc}}$ within an Arrhenius representation, revealing clear non-Arrhenius behavior. Figure S\ref{fig:resistance_lunkenheimer}(c) shows a representation that should lead to a linearization of the data for variable range hopping (VRH) \cite{Mott1979}, an often-found transport mechanism for localized electronic charge carriers at low temperatures. Again, no linear region can be identified. The same is the case for two or one-dimensional VRH, where the predicted exponent 1/4 is replaced by 1/3 or 1/2, respectively. Anyway, the significant reduction of temperature dependence revealed by these figures at low temperatures may well indicate a crossover to electronic tunneling transport.

\bibliography{k-BETSMn_paper_preprint-combined_v2}

\begin{thebibliography}{67}%
\makeatletter
\providecommand \@ifxundefined [1]{%
 \@ifx{#1\undefined}
}%
\providecommand \@ifnum [1]{%
 \ifnum #1\expandafter \@firstoftwo
 \else \expandafter \@secondoftwo
 \fi
}%
\providecommand \@ifx [1]{%
 \ifx #1\expandafter \@firstoftwo
 \else \expandafter \@secondoftwo
 \fi
}%
\providecommand \natexlab [1]{#1}%
\providecommand \enquote  [1]{``#1''}%
\providecommand \bibnamefont  [1]{#1}%
\providecommand \bibfnamefont [1]{#1}%
\providecommand \citenamefont [1]{#1}%
\providecommand \href@noop [0]{\@secondoftwo}%
\providecommand \href [0]{\begingroup \@sanitize@url \@href}%
\providecommand \@href[1]{\@@startlink{#1}\@@href}%
\providecommand \@@href[1]{\endgroup#1\@@endlink}%
\providecommand \@sanitize@url [0]{\catcode `\\12\catcode `\$12\catcode
  `\&12\catcode `\#12\catcode `\^12\catcode `\_12\catcode `\%12\relax}%
\providecommand \@@startlink[1]{}%
\providecommand \@@endlink[0]{}%
\providecommand \url  [0]{\begingroup\@sanitize@url \@url }%
\providecommand \@url [1]{\endgroup\@href {#1}{\urlprefix }}%
\providecommand \urlprefix  [0]{URL }%
\providecommand \Eprint [0]{\href }%
\providecommand \doibase [0]{http://dx.doi.org/}%
\providecommand \selectlanguage [0]{\@gobble}%
\providecommand \bibinfo  [0]{\@secondoftwo}%
\providecommand \bibfield  [0]{\@secondoftwo}%
\providecommand \translation [1]{[#1]}%
\providecommand \BibitemOpen [0]{}%
\providecommand \bibitemStop [0]{}%
\providecommand \bibitemNoStop [0]{.\EOS\space}%
\providecommand \EOS [0]{\spacefactor3000\relax}%
\providecommand \BibitemShut  [1]{\csname bibitem#1\endcsname}%
\let\auto@bib@innerbib\@empty
\bibitem [{\citenamefont {van~den Brink}\ and\ \citenamefont
  {Khomskii}(2008)}]{Brink2008}%
  \BibitemOpen
  \bibfield  {author} {\bibinfo {author} {\bibfnamefont {J.}~\bibnamefont
  {van~den Brink}}\ and\ \bibinfo {author} {\bibfnamefont {D.~I.}\ \bibnamefont
  {Khomskii}},\ }\href {http://stacks.iop.org/0953-8984/20/i=43/a=434217}
  {\bibfield  {journal} {\bibinfo  {journal} {Journal of Physics: Condensed
  Matter}\ }\textbf {\bibinfo {volume} {20}},\ \bibinfo {pages} {434217}
  (\bibinfo {year} {2008})}\BibitemShut {NoStop}%
\bibitem [{\citenamefont {Horiuchi}\ and\ \citenamefont
  {Tokura}(2008)}]{Horiuchi2008}%
  \BibitemOpen
  \bibfield  {author} {\bibinfo {author} {\bibfnamefont {S.}~\bibnamefont
  {Horiuchi}}\ and\ \bibinfo {author} {\bibfnamefont {Y.}~\bibnamefont
  {Tokura}},\ }\href@noop {} {\bibfield  {journal} {\bibinfo  {journal} {Nature
  Materials}\ }\textbf {\bibinfo {volume} {7}},\ \bibinfo {pages} {357}
  (\bibinfo {year} {2008})}\BibitemShut {NoStop}%
\bibitem [{\citenamefont {Ishihara}(2010)}]{Ishihara2010}%
  \BibitemOpen
  \bibfield  {author} {\bibinfo {author} {\bibfnamefont {S.}~\bibnamefont
  {Ishihara}},\ }\href {https://doi.org/10.1143/JPSJ.79.011010} {\bibfield
  {journal} {\bibinfo  {journal} {Journal of the Physical Society of Japan}\
  }\textbf {\bibinfo {volume} {79}},\ \bibinfo {pages} {011010} (\bibinfo
  {year} {2010})},\ \Eprint
  {http://arxiv.org/abs/https://doi.org/10.1143/JPSJ.79.011010}
  {https://doi.org/10.1143/JPSJ.79.011010} \BibitemShut {NoStop}%
\bibitem [{\citenamefont {Kimura}\ \emph {et~al.}(2003)\citenamefont {Kimura},
  \citenamefont {Goto}, \citenamefont {Shintani}, \citenamefont {Ishizaka},
  \citenamefont {Arima},\ and\ \citenamefont {Tokura}}]{Kimura2003}%
  \BibitemOpen
  \bibfield  {author} {\bibinfo {author} {\bibfnamefont {T.}~\bibnamefont
  {Kimura}}, \bibinfo {author} {\bibfnamefont {T.}~\bibnamefont {Goto}},
  \bibinfo {author} {\bibfnamefont {H.}~\bibnamefont {Shintani}}, \bibinfo
  {author} {\bibfnamefont {K.}~\bibnamefont {Ishizaka}}, \bibinfo {author}
  {\bibfnamefont {T.}~\bibnamefont {Arima}}, \ and\ \bibinfo {author}
  {\bibfnamefont {Y.}~\bibnamefont {Tokura}},\ }\href@noop {} {\bibfield
  {journal} {\bibinfo  {journal} {Nature}\ }\textbf {\bibinfo {volume} {426}},\
  \bibinfo {pages} {55} (\bibinfo {year} {2003})}\BibitemShut {NoStop}%
\bibitem [{\citenamefont {Lunkenheimer}\ and\ \citenamefont
  {Loidl}(2015)}]{Lunkenheimer2015a}%
  \BibitemOpen
  \bibfield  {author} {\bibinfo {author} {\bibfnamefont {P.}~\bibnamefont
  {Lunkenheimer}}\ and\ \bibinfo {author} {\bibfnamefont {A.}~\bibnamefont
  {Loidl}},\ }\href {\doibase 10.1088/0953-8984/27/37/373001} {\bibfield
  {journal} {\bibinfo  {journal} {Journal of Physics: Condensed Matter}\
  }\textbf {\bibinfo {volume} {27}},\ \bibinfo {pages} {373001} (\bibinfo
  {year} {2015})}\BibitemShut {NoStop}%
\bibitem [{\citenamefont {Tomi{\'{c}}}\ and\ \citenamefont
  {Dressel}(2015)}]{Tomic2015}%
  \BibitemOpen
  \bibfield  {author} {\bibinfo {author} {\bibfnamefont {S.}~\bibnamefont
  {Tomi{\'{c}}}}\ and\ \bibinfo {author} {\bibfnamefont {M.}~\bibnamefont
  {Dressel}},\ }\href {\doibase 10.1088/0034-4885/78/9/096501} {\bibfield
  {journal} {\bibinfo  {journal} {Rep. Prog. Phys.}\ }\textbf {\bibinfo
  {volume} {78}},\ \bibinfo {pages} {096501} (\bibinfo {year}
  {2015})}\BibitemShut {NoStop}%
\bibitem [{\citenamefont {Naka}\ and\ \citenamefont
  {Ishihara}(2010)}]{Naka2010}%
  \BibitemOpen
  \bibfield  {author} {\bibinfo {author} {\bibfnamefont {M.}~\bibnamefont
  {Naka}}\ and\ \bibinfo {author} {\bibfnamefont {S.}~\bibnamefont
  {Ishihara}},\ }\href {\doibase 10.1143/jpsj.79.063707} {\bibfield  {journal}
  {\bibinfo  {journal} {J. Phys. Soc. Jpn.}\ }\textbf {\bibinfo {volume}
  {79}},\ \bibinfo {pages} {063707} (\bibinfo {year} {2010})}\BibitemShut
  {NoStop}%
\bibitem [{\citenamefont {Ishihara}(2014)}]{Ishihara2014}%
  \BibitemOpen
  \bibfield  {author} {\bibinfo {author} {\bibfnamefont {S.}~\bibnamefont
  {Ishihara}},\ }\href {\doibase 10.1088/0953-8984/26/49/493201} {\bibfield
  {journal} {\bibinfo  {journal} {J. Phys.: Condens. Matter}\ }\textbf
  {\bibinfo {volume} {26}},\ \bibinfo {pages} {493201} (\bibinfo {year}
  {2014})}\BibitemShut {NoStop}%
\bibitem [{\citenamefont {Lunkenheimer}\ \emph {et~al.}(2012)\citenamefont
  {Lunkenheimer}, \citenamefont {M\"{u}ller}, \citenamefont {Krohns},
  \citenamefont {Schrettle}, \citenamefont {Loidl}, \citenamefont {Hartmann},
  \citenamefont {Rommel}, \citenamefont {de~Souza}, \citenamefont {Hotta},
  \citenamefont {Schlueter},\ and\ \citenamefont {Lang}}]{Lunkenheimer2012}%
  \BibitemOpen
  \bibfield  {author} {\bibinfo {author} {\bibfnamefont {P.}~\bibnamefont
  {Lunkenheimer}}, \bibinfo {author} {\bibfnamefont {J.}~\bibnamefont
  {M\"{u}ller}}, \bibinfo {author} {\bibfnamefont {S.}~\bibnamefont {Krohns}},
  \bibinfo {author} {\bibfnamefont {F.}~\bibnamefont {Schrettle}}, \bibinfo
  {author} {\bibfnamefont {A.}~\bibnamefont {Loidl}}, \bibinfo {author}
  {\bibfnamefont {B.}~\bibnamefont {Hartmann}}, \bibinfo {author}
  {\bibfnamefont {R.}~\bibnamefont {Rommel}}, \bibinfo {author} {\bibfnamefont
  {M.}~\bibnamefont {de~Souza}}, \bibinfo {author} {\bibfnamefont
  {C.}~\bibnamefont {Hotta}}, \bibinfo {author} {\bibfnamefont {J.~A.}\
  \bibnamefont {Schlueter}}, \ and\ \bibinfo {author} {\bibfnamefont
  {M.}~\bibnamefont {Lang}},\ }\href {\doibase 10.1038/nmat3400} {\bibfield
  {journal} {\bibinfo  {journal} {Nature Materials}\ }\textbf {\bibinfo
  {volume} {11}},\ \bibinfo {pages} {755} (\bibinfo {year} {2012})}\BibitemShut
  {NoStop}%
\bibitem [{\citenamefont {Drichko}\ \emph {et~al.}(2014)\citenamefont
  {Drichko}, \citenamefont {Beyer}, \citenamefont {Rose}, \citenamefont
  {Dressel}, \citenamefont {Schlueter}, \citenamefont {Turunova}, \citenamefont
  {Zhilyaeva},\ and\ \citenamefont {Lyubovskaya}}]{Drichko2014}%
  \BibitemOpen
  \bibfield  {author} {\bibinfo {author} {\bibfnamefont {N.}~\bibnamefont
  {Drichko}}, \bibinfo {author} {\bibfnamefont {R.}~\bibnamefont {Beyer}},
  \bibinfo {author} {\bibfnamefont {E.}~\bibnamefont {Rose}}, \bibinfo {author}
  {\bibfnamefont {M.}~\bibnamefont {Dressel}}, \bibinfo {author} {\bibfnamefont
  {J.~A.}\ \bibnamefont {Schlueter}}, \bibinfo {author} {\bibfnamefont {S.~A.}\
  \bibnamefont {Turunova}}, \bibinfo {author} {\bibfnamefont {E.~I.}\
  \bibnamefont {Zhilyaeva}}, \ and\ \bibinfo {author} {\bibfnamefont {R.~N.}\
  \bibnamefont {Lyubovskaya}},\ }\href {\doibase 10.1103/PhysRevB.89.075133}
  {\bibfield  {journal} {\bibinfo  {journal} {Phys. Rev. B}\ }\textbf {\bibinfo
  {volume} {89}},\ \bibinfo {pages} {075133} (\bibinfo {year}
  {2014})}\BibitemShut {NoStop}%
\bibitem [{\citenamefont {Gati}\ \emph
  {et~al.}(2018{\natexlab{a}})\citenamefont {Gati}, \citenamefont {Fischer},
  \citenamefont {Lunkenheimer}, \citenamefont {Zielke}, \citenamefont
  {K\"ohler}, \citenamefont {Kolb}, \citenamefont {von Nidda}, \citenamefont
  {Winter}, \citenamefont {Schubert}, \citenamefont {Schlueter}, \citenamefont
  {Jeschke}, \citenamefont {Valent\'{\i}},\ and\ \citenamefont
  {Lang}}]{Gati2018b}%
  \BibitemOpen
  \bibfield  {author} {\bibinfo {author} {\bibfnamefont {E.}~\bibnamefont
  {Gati}}, \bibinfo {author} {\bibfnamefont {J.~K.~H.}\ \bibnamefont
  {Fischer}}, \bibinfo {author} {\bibfnamefont {P.}~\bibnamefont
  {Lunkenheimer}}, \bibinfo {author} {\bibfnamefont {D.}~\bibnamefont
  {Zielke}}, \bibinfo {author} {\bibfnamefont {S.}~\bibnamefont {K\"ohler}},
  \bibinfo {author} {\bibfnamefont {F.}~\bibnamefont {Kolb}}, \bibinfo {author}
  {\bibfnamefont {H.-A.~K.}\ \bibnamefont {von Nidda}}, \bibinfo {author}
  {\bibfnamefont {S.~M.}\ \bibnamefont {Winter}}, \bibinfo {author}
  {\bibfnamefont {H.}~\bibnamefont {Schubert}}, \bibinfo {author}
  {\bibfnamefont {J.~A.}\ \bibnamefont {Schlueter}}, \bibinfo {author}
  {\bibfnamefont {H.~O.}\ \bibnamefont {Jeschke}}, \bibinfo {author}
  {\bibfnamefont {R.}~\bibnamefont {Valent\'{\i}}}, \ and\ \bibinfo {author}
  {\bibfnamefont {M.}~\bibnamefont {Lang}},\ }\href {\doibase
  10.1103/PhysRevLett.120.247601} {\bibfield  {journal} {\bibinfo  {journal}
  {Phys. Rev. Lett.}\ }\textbf {\bibinfo {volume} {120}},\ \bibinfo {pages}
  {247601} (\bibinfo {year} {2018}{\natexlab{a}})}\BibitemShut {NoStop}%
\bibitem [{\citenamefont {Gati}\ \emph
  {et~al.}(2018{\natexlab{b}})\citenamefont {Gati}, \citenamefont {Winter},
  \citenamefont {Schlueter}, \citenamefont {Schubert}, \citenamefont
  {M{\"{u}}ller},\ and\ \citenamefont {Lang}}]{Gati2018}%
  \BibitemOpen
  \bibfield  {author} {\bibinfo {author} {\bibfnamefont {E.}~\bibnamefont
  {Gati}}, \bibinfo {author} {\bibfnamefont {S.~M.}\ \bibnamefont {Winter}},
  \bibinfo {author} {\bibfnamefont {J.~A.}\ \bibnamefont {Schlueter}}, \bibinfo
  {author} {\bibfnamefont {H.}~\bibnamefont {Schubert}}, \bibinfo {author}
  {\bibfnamefont {J.}~\bibnamefont {M{\"{u}}ller}}, \ and\ \bibinfo {author}
  {\bibfnamefont {M.}~\bibnamefont {Lang}},\ }\href {\doibase
  10.1103/PhysRevB.97.075115} {\bibfield  {journal} {\bibinfo  {journal}
  {Physical Review B}\ }\textbf {\bibinfo {volume} {97}},\ \bibinfo {pages}
  {075115} (\bibinfo {year} {2018}{\natexlab{b}})}\BibitemShut {NoStop}%
\bibitem [{\citenamefont {Abdel-Jawad}\ \emph {et~al.}(2010)\citenamefont
  {Abdel-Jawad}, \citenamefont {Terasaki}, \citenamefont {Sasaki},
  \citenamefont {Yoneyama}, \citenamefont {Kobayashi}, \citenamefont {Uesu},\
  and\ \citenamefont {Hotta}}]{Abdel-Jawad2010}%
  \BibitemOpen
  \bibfield  {author} {\bibinfo {author} {\bibfnamefont {M.}~\bibnamefont
  {Abdel-Jawad}}, \bibinfo {author} {\bibfnamefont {I.}~\bibnamefont
  {Terasaki}}, \bibinfo {author} {\bibfnamefont {T.}~\bibnamefont {Sasaki}},
  \bibinfo {author} {\bibfnamefont {N.}~\bibnamefont {Yoneyama}}, \bibinfo
  {author} {\bibfnamefont {N.}~\bibnamefont {Kobayashi}}, \bibinfo {author}
  {\bibfnamefont {Y.}~\bibnamefont {Uesu}}, \ and\ \bibinfo {author}
  {\bibfnamefont {C.}~\bibnamefont {Hotta}},\ }\href {\doibase
  10.1103/physrevb.82.125119} {\bibfield  {journal} {\bibinfo  {journal} {Phys.
  Rev. B}\ }\textbf {\bibinfo {volume} {82}},\ \bibinfo {pages} {125119}
  (\bibinfo {year} {2010})}\BibitemShut {NoStop}%
\bibitem [{\citenamefont {Pinteri\ifmmode~\acute{c}\else \'{c}\fi{}}\ \emph
  {et~al.}(2016)\citenamefont {Pinteri\ifmmode~\acute{c}\else \'{c}\fi{}},
  \citenamefont {Lazi\ifmmode~\acute{c}\else \'{c}\fi{}}, \citenamefont
  {Pustogow}, \citenamefont {Ivek}, \citenamefont {Kuve\ifmmode \check{z}\else
  \v{z}\fi{}di\ifmmode~\acute{c}\else \'{c}\fi{}}, \citenamefont {Milat},
  \citenamefont {Gumhalter}, \citenamefont {Basleti\ifmmode~\acute{c}\else
  \'{c}\fi{}}, \citenamefont {\ifmmode~\check{C}\else \v{C}\fi{}ulo},
  \citenamefont {Korin-Hamzi\ifmmode~\acute{c}\else \'{c}\fi{}}, \citenamefont
  {L\"ohle}, \citenamefont {H\"ubner}, \citenamefont {Sanz~Alonso},
  \citenamefont {Hiramatsu}, \citenamefont {Yoshida}, \citenamefont {Saito},
  \citenamefont {Dressel},\ and\ \citenamefont {Tomi\ifmmode~\acute{c}\else
  \'{c}\fi{}}}]{Pinteric2016}%
  \BibitemOpen
  \bibfield  {author} {\bibinfo {author} {\bibfnamefont {M.}~\bibnamefont
  {Pinteri\ifmmode~\acute{c}\else \'{c}\fi{}}}, \bibinfo {author}
  {\bibfnamefont {P.}~\bibnamefont {Lazi\ifmmode~\acute{c}\else \'{c}\fi{}}},
  \bibinfo {author} {\bibfnamefont {A.}~\bibnamefont {Pustogow}}, \bibinfo
  {author} {\bibfnamefont {T.}~\bibnamefont {Ivek}}, \bibinfo {author}
  {\bibfnamefont {M.}~\bibnamefont {Kuve\ifmmode \check{z}\else
  \v{z}\fi{}di\ifmmode~\acute{c}\else \'{c}\fi{}}}, \bibinfo {author}
  {\bibfnamefont {O.}~\bibnamefont {Milat}}, \bibinfo {author} {\bibfnamefont
  {B.}~\bibnamefont {Gumhalter}}, \bibinfo {author} {\bibfnamefont
  {M.}~\bibnamefont {Basleti\ifmmode~\acute{c}\else \'{c}\fi{}}}, \bibinfo
  {author} {\bibfnamefont {M.}~\bibnamefont {\ifmmode~\check{C}\else
  \v{C}\fi{}ulo}}, \bibinfo {author} {\bibfnamefont {B.}~\bibnamefont
  {Korin-Hamzi\ifmmode~\acute{c}\else \'{c}\fi{}}}, \bibinfo {author}
  {\bibfnamefont {A.}~\bibnamefont {L\"ohle}}, \bibinfo {author} {\bibfnamefont
  {R.}~\bibnamefont {H\"ubner}}, \bibinfo {author} {\bibfnamefont
  {M.}~\bibnamefont {Sanz~Alonso}}, \bibinfo {author} {\bibfnamefont
  {T.}~\bibnamefont {Hiramatsu}}, \bibinfo {author} {\bibfnamefont
  {Y.}~\bibnamefont {Yoshida}}, \bibinfo {author} {\bibfnamefont
  {G.}~\bibnamefont {Saito}}, \bibinfo {author} {\bibfnamefont
  {M.}~\bibnamefont {Dressel}}, \ and\ \bibinfo {author} {\bibfnamefont
  {S.}~\bibnamefont {Tomi\ifmmode~\acute{c}\else \'{c}\fi{}}},\ }\href
  {\doibase 10.1103/PhysRevB.94.161105} {\bibfield  {journal} {\bibinfo
  {journal} {Phys. Rev. B}\ }\textbf {\bibinfo {volume} {94}},\ \bibinfo
  {pages} {161105} (\bibinfo {year} {2016})}\BibitemShut {NoStop}%
\bibitem [{\citenamefont {Iguchi}\ \emph {et~al.}(2013)\citenamefont {Iguchi},
  \citenamefont {Sasaki}, \citenamefont {Yoneyama}, \citenamefont {Taniguchi},
  \citenamefont {Nishizaki},\ and\ \citenamefont {Sasaki}}]{Iguchi2013}%
  \BibitemOpen
  \bibfield  {author} {\bibinfo {author} {\bibfnamefont {S.}~\bibnamefont
  {Iguchi}}, \bibinfo {author} {\bibfnamefont {S.}~\bibnamefont {Sasaki}},
  \bibinfo {author} {\bibfnamefont {N.}~\bibnamefont {Yoneyama}}, \bibinfo
  {author} {\bibfnamefont {H.}~\bibnamefont {Taniguchi}}, \bibinfo {author}
  {\bibfnamefont {T.}~\bibnamefont {Nishizaki}}, \ and\ \bibinfo {author}
  {\bibfnamefont {T.}~\bibnamefont {Sasaki}},\ }\href@noop {} {\bibfield
  {journal} {\bibinfo  {journal} {Phys. Rev. B}\ }\textbf {\bibinfo {volume}
  {87}},\ \bibinfo {pages} {075107} (\bibinfo {year} {2013})}\BibitemShut
  {NoStop}%
\bibitem [{\citenamefont {Lunkenheimer}\ \emph {et~al.}(2015)\citenamefont
  {Lunkenheimer}, \citenamefont {Hartmann}, \citenamefont {Lang}, \citenamefont
  {M\"uller}, \citenamefont {Schweitzer}, \citenamefont {Krohns},\ and\
  \citenamefont {Loidl}}]{Lunkenheimer2015}%
  \BibitemOpen
  \bibfield  {author} {\bibinfo {author} {\bibfnamefont {P.}~\bibnamefont
  {Lunkenheimer}}, \bibinfo {author} {\bibfnamefont {B.}~\bibnamefont
  {Hartmann}}, \bibinfo {author} {\bibfnamefont {M.}~\bibnamefont {Lang}},
  \bibinfo {author} {\bibfnamefont {J.}~\bibnamefont {M\"uller}}, \bibinfo
  {author} {\bibfnamefont {D.}~\bibnamefont {Schweitzer}}, \bibinfo {author}
  {\bibfnamefont {S.}~\bibnamefont {Krohns}}, \ and\ \bibinfo {author}
  {\bibfnamefont {A.}~\bibnamefont {Loidl}},\ }\href {\doibase
  10.1103/PhysRevB.91.245132} {\bibfield  {journal} {\bibinfo  {journal} {Phys.
  Rev. B}\ }\textbf {\bibinfo {volume} {91}},\ \bibinfo {pages} {245132}
  (\bibinfo {year} {2015})}\BibitemShut {NoStop}%
\bibitem [{\citenamefont {Abdel-Jawad}\ \emph {et~al.}(2013)\citenamefont
  {Abdel-Jawad}, \citenamefont {Tajima}, \citenamefont {Kato},\ and\
  \citenamefont {Terasaki}}]{Abdel-Jawad2013}%
  \BibitemOpen
  \bibfield  {author} {\bibinfo {author} {\bibfnamefont {M.}~\bibnamefont
  {Abdel-Jawad}}, \bibinfo {author} {\bibfnamefont {N.}~\bibnamefont {Tajima}},
  \bibinfo {author} {\bibfnamefont {R.}~\bibnamefont {Kato}}, \ and\ \bibinfo
  {author} {\bibfnamefont {I.}~\bibnamefont {Terasaki}},\ }\href {\doibase
  10.1103/physrevb.88.075139} {\bibfield  {journal} {\bibinfo  {journal} {Phys.
  Rev. B}\ }\textbf {\bibinfo {volume} {88}} (\bibinfo {year} {2013}),\
  10.1103/physrevb.88.075139}\BibitemShut {NoStop}%
\bibitem [{\citenamefont {Hotta}(2010)}]{Hotta2010}%
  \BibitemOpen
  \bibfield  {author} {\bibinfo {author} {\bibfnamefont {C.}~\bibnamefont
  {Hotta}},\ }\href@noop {} {\bibfield  {journal} {\bibinfo  {journal} {Phys.
  Rev. B}\ }\textbf {\bibinfo {volume} {82}},\ \bibinfo {pages} {241104(R)}
  (\bibinfo {year} {2010})}\BibitemShut {NoStop}%
\bibitem [{\citenamefont {Deglint}\ \emph {et~al.}(2022)\citenamefont
  {Deglint}, \citenamefont {Akella},\ and\ \citenamefont
  {Kennett}}]{Deglint2022}%
  \BibitemOpen
  \bibfield  {author} {\bibinfo {author} {\bibfnamefont {M.~B.}\ \bibnamefont
  {Deglint}}, \bibinfo {author} {\bibfnamefont {K.}~\bibnamefont {Akella}}, \
  and\ \bibinfo {author} {\bibfnamefont {M.~P.}\ \bibnamefont {Kennett}},\
  }\href {\doibase 10.1103/PhysRevB.106.085123} {\bibfield  {journal} {\bibinfo
   {journal} {Phys. Rev. B}\ }\textbf {\bibinfo {volume} {106}},\ \bibinfo
  {pages} {085123} (\bibinfo {year} {2022})}\BibitemShut {NoStop}%
\bibitem [{\citenamefont {Bokov}\ and\ \citenamefont {Ye}(2006)}]{Bokov2006}%
  \BibitemOpen
  \bibfield  {author} {\bibinfo {author} {\bibfnamefont {A.~A.}\ \bibnamefont
  {Bokov}}\ and\ \bibinfo {author} {\bibfnamefont {Z.~G.}\ \bibnamefont {Ye}},\
  }\href {\doibase 10.1007/s10853-005-5915-7} {\bibfield  {journal} {\bibinfo
  {journal} {Journal of Materials Science}\ }\textbf {\bibinfo {volume} {41}},\
  \bibinfo {pages} {31} (\bibinfo {year} {2006})}\BibitemShut {NoStop}%
\bibitem [{\citenamefont {Fu}\ \emph {et~al.}(2009)\citenamefont {Fu},
  \citenamefont {Taniguchi}, \citenamefont {Itoh}, \citenamefont
  {ya~Koshihara}, \citenamefont {Yamamoto},\ and\ \citenamefont
  {Mori}}]{Fu2009}%
  \BibitemOpen
  \bibfield  {author} {\bibinfo {author} {\bibfnamefont {D.}~\bibnamefont
  {Fu}}, \bibinfo {author} {\bibfnamefont {H.}~\bibnamefont {Taniguchi}},
  \bibinfo {author} {\bibfnamefont {M.}~\bibnamefont {Itoh}}, \bibinfo {author}
  {\bibfnamefont {S.}~\bibnamefont {ya~Koshihara}}, \bibinfo {author}
  {\bibfnamefont {N.}~\bibnamefont {Yamamoto}}, \ and\ \bibinfo {author}
  {\bibfnamefont {S.}~\bibnamefont {Mori}},\ }\href {\doibase
  10.1103/physrevlett.103.207601} {\bibfield  {journal} {\bibinfo  {journal}
  {Phys. Rev. Lett.}\ }\textbf {\bibinfo {volume} {103}},\ \bibinfo {pages}
  {207601} (\bibinfo {year} {2009})}\BibitemShut {NoStop}%
\bibitem [{\citenamefont {Kushch}\ \emph {et~al.}(2008)\citenamefont {Kushch},
  \citenamefont {Yagubskii}, \citenamefont {Kartsovnik}, \citenamefont
  {Buravov}, \citenamefont {Dubrovskii}, \citenamefont {Chekhlov},\ and\
  \citenamefont {Biberacher}}]{Kushch2008}%
  \BibitemOpen
  \bibfield  {author} {\bibinfo {author} {\bibfnamefont {N.~D.}\ \bibnamefont
  {Kushch}}, \bibinfo {author} {\bibfnamefont {E.~B.}\ \bibnamefont
  {Yagubskii}}, \bibinfo {author} {\bibfnamefont {M.~V.}\ \bibnamefont
  {Kartsovnik}}, \bibinfo {author} {\bibfnamefont {L.~I.}\ \bibnamefont
  {Buravov}}, \bibinfo {author} {\bibfnamefont {A.~D.}\ \bibnamefont
  {Dubrovskii}}, \bibinfo {author} {\bibfnamefont {A.~N.}\ \bibnamefont
  {Chekhlov}}, \ and\ \bibinfo {author} {\bibfnamefont {W.}~\bibnamefont
  {Biberacher}},\ }\href@noop {} {\bibfield  {journal} {\bibinfo  {journal}
  {Journal of the American Chemical Society}\ }\textbf {\bibinfo {volume}
  {130}},\ \bibinfo {pages} {7238} (\bibinfo {year} {2008})}\BibitemShut
  {NoStop}%
\bibitem [{\citenamefont {Zverev}\ \emph {et~al.}(2010)\citenamefont {Zverev},
  \citenamefont {Kartsovnik}, \citenamefont {Biberacher}, \citenamefont
  {Khasanov}, \citenamefont {Shibaeva}, \citenamefont {Ouahab}, \citenamefont
  {Toupet}, \citenamefont {Kushch}, \citenamefont {Yagubskii},\ and\
  \citenamefont {Canadell}}]{Zverev2010}%
  \BibitemOpen
  \bibfield  {author} {\bibinfo {author} {\bibfnamefont {V.~N.}\ \bibnamefont
  {Zverev}}, \bibinfo {author} {\bibfnamefont {M.~V.}\ \bibnamefont
  {Kartsovnik}}, \bibinfo {author} {\bibfnamefont {W.}~\bibnamefont
  {Biberacher}}, \bibinfo {author} {\bibfnamefont {S.~S.}\ \bibnamefont
  {Khasanov}}, \bibinfo {author} {\bibfnamefont {R.~P.}\ \bibnamefont
  {Shibaeva}}, \bibinfo {author} {\bibfnamefont {L.}~\bibnamefont {Ouahab}},
  \bibinfo {author} {\bibfnamefont {L.}~\bibnamefont {Toupet}}, \bibinfo
  {author} {\bibfnamefont {N.~D.}\ \bibnamefont {Kushch}}, \bibinfo {author}
  {\bibfnamefont {E.~B.}\ \bibnamefont {Yagubskii}}, \ and\ \bibinfo {author}
  {\bibfnamefont {E.}~\bibnamefont {Canadell}},\ }\href {\doibase
  10.1103/PhysRevB.82.155123} {\bibfield  {journal} {\bibinfo  {journal} {Phys.
  Rev. B}\ }\textbf {\bibinfo {volume} {82}},\ \bibinfo {pages} {155123}
  (\bibinfo {year} {2010})}\BibitemShut {NoStop}%
\bibitem [{\citenamefont {Uji}\ \emph {et~al.}(2001)\citenamefont {Uji},
  \citenamefont {Shinagawa}, \citenamefont {Terashima}, \citenamefont {Yakabe},
  \citenamefont {Terai}, \citenamefont {Tokumoto}, \citenamefont {Kobayashi},
  \citenamefont {Tanaka},\ and\ \citenamefont {Kobayashi}}]{Uji2001}%
  \BibitemOpen
  \bibfield  {author} {\bibinfo {author} {\bibfnamefont {S.}~\bibnamefont
  {Uji}}, \bibinfo {author} {\bibfnamefont {H.}~\bibnamefont {Shinagawa}},
  \bibinfo {author} {\bibfnamefont {T.}~\bibnamefont {Terashima}}, \bibinfo
  {author} {\bibfnamefont {T.}~\bibnamefont {Yakabe}}, \bibinfo {author}
  {\bibfnamefont {Y.}~\bibnamefont {Terai}}, \bibinfo {author} {\bibfnamefont
  {M.}~\bibnamefont {Tokumoto}}, \bibinfo {author} {\bibfnamefont
  {A.}~\bibnamefont {Kobayashi}}, \bibinfo {author} {\bibfnamefont
  {H.}~\bibnamefont {Tanaka}}, \ and\ \bibinfo {author} {\bibfnamefont
  {H.}~\bibnamefont {Kobayashi}},\ }\href {\doibase 10.1038/35073531}
  {\bibfield  {journal} {\bibinfo  {journal} {Nature}\ }\textbf {\bibinfo
  {volume} {410}},\ \bibinfo {pages} {908} (\bibinfo {year}
  {2001})}\BibitemShut {NoStop}%
\bibitem [{\citenamefont {Fujiwara}\ \emph {et~al.}(2002)\citenamefont
  {Fujiwara}, \citenamefont {Kobayashi}, \citenamefont {Fujiwara},\ and\
  \citenamefont {Kobayashi}}]{Fujiwara2002}%
  \BibitemOpen
  \bibfield  {author} {\bibinfo {author} {\bibfnamefont {H.}~\bibnamefont
  {Fujiwara}}, \bibinfo {author} {\bibfnamefont {H.}~\bibnamefont {Kobayashi}},
  \bibinfo {author} {\bibfnamefont {E.}~\bibnamefont {Fujiwara}}, \ and\
  \bibinfo {author} {\bibfnamefont {A.}~\bibnamefont {Kobayashi}},\ }\href
  {\doibase 10.1021/ja026067z} {\bibfield  {journal} {\bibinfo  {journal}
  {Journal of the American Chemical Society}\ }\textbf {\bibinfo {volume}
  {124}},\ \bibinfo {pages} {6816} (\bibinfo {year} {2002})}\BibitemShut
  {NoStop}%
\bibitem [{\citenamefont {Riedl}\ \emph {et~al.}(2021)\citenamefont {Riedl},
  \citenamefont {Gati}, \citenamefont {Zielke}, \citenamefont {Hartmann},
  \citenamefont {Vyaselev}, \citenamefont {Kushch}, \citenamefont {Jeschke},
  \citenamefont {Lang}, \citenamefont {Valent\'{\i}}, \citenamefont
  {Kartsovnik},\ and\ \citenamefont {Winter}}]{Riedl2021}%
  \BibitemOpen
  \bibfield  {author} {\bibinfo {author} {\bibfnamefont {K.}~\bibnamefont
  {Riedl}}, \bibinfo {author} {\bibfnamefont {E.}~\bibnamefont {Gati}},
  \bibinfo {author} {\bibfnamefont {D.}~\bibnamefont {Zielke}}, \bibinfo
  {author} {\bibfnamefont {S.}~\bibnamefont {Hartmann}}, \bibinfo {author}
  {\bibfnamefont {O.~M.}\ \bibnamefont {Vyaselev}}, \bibinfo {author}
  {\bibfnamefont {N.~D.}\ \bibnamefont {Kushch}}, \bibinfo {author}
  {\bibfnamefont {H.~O.}\ \bibnamefont {Jeschke}}, \bibinfo {author}
  {\bibfnamefont {M.}~\bibnamefont {Lang}}, \bibinfo {author} {\bibfnamefont
  {R.}~\bibnamefont {Valent\'{\i}}}, \bibinfo {author} {\bibfnamefont {M.~V.}\
  \bibnamefont {Kartsovnik}}, \ and\ \bibinfo {author} {\bibfnamefont {S.~M.}\
  \bibnamefont {Winter}},\ }\href {\doibase 10.1103/PhysRevLett.127.147204}
  {\bibfield  {journal} {\bibinfo  {journal} {Phys. Rev. Lett.}\ }\textbf
  {\bibinfo {volume} {127}},\ \bibinfo {pages} {147204} (\bibinfo {year}
  {2021})}\BibitemShut {NoStop}%
\bibitem [{\citenamefont {Vyaselev}\ \emph {et~al.}(2017)\citenamefont
  {Vyaselev}, \citenamefont {Biberacher}, \citenamefont {Kushch},\ and\
  \citenamefont {Kartsovnik}}]{Vyaselev2017}%
  \BibitemOpen
  \bibfield  {author} {\bibinfo {author} {\bibfnamefont {O.~M.}\ \bibnamefont
  {Vyaselev}}, \bibinfo {author} {\bibfnamefont {W.}~\bibnamefont
  {Biberacher}}, \bibinfo {author} {\bibfnamefont {N.~D.}\ \bibnamefont
  {Kushch}}, \ and\ \bibinfo {author} {\bibfnamefont {M.~V.}\ \bibnamefont
  {Kartsovnik}},\ }\href {\doibase 10.1103/PhysRevB.96.205154} {\bibfield
  {journal} {\bibinfo  {journal} {Phys. Rev. B}\ }\textbf {\bibinfo {volume}
  {96}},\ \bibinfo {pages} {205154} (\bibinfo {year} {2017})}\BibitemShut
  {NoStop}%
\bibitem [{\citenamefont {Kartsovnik}\ \emph {et~al.}(2017)\citenamefont
  {Kartsovnik}, \citenamefont {Zverev}, \citenamefont {Biberacher},
  \citenamefont {Simonov}, \citenamefont {Sheikin}, \citenamefont {Kushch},\
  and\ \citenamefont {Yagubskii}}]{Kartsovnik2017}%
  \BibitemOpen
  \bibfield  {author} {\bibinfo {author} {\bibfnamefont {M.~V.}\ \bibnamefont
  {Kartsovnik}}, \bibinfo {author} {\bibfnamefont {V.~N.}\ \bibnamefont
  {Zverev}}, \bibinfo {author} {\bibfnamefont {W.}~\bibnamefont {Biberacher}},
  \bibinfo {author} {\bibfnamefont {S.~V.}\ \bibnamefont {Simonov}}, \bibinfo
  {author} {\bibfnamefont {I.}~\bibnamefont {Sheikin}}, \bibinfo {author}
  {\bibfnamefont {N.~D.}\ \bibnamefont {Kushch}}, \ and\ \bibinfo {author}
  {\bibfnamefont {E.~B.}\ \bibnamefont {Yagubskii}},\ }\href {\doibase
  10.1063/1.4976634} {\bibfield  {journal} {\bibinfo  {journal} {Low
  Temperature Physics}\ }\textbf {\bibinfo {volume} {43}},\ \bibinfo {pages}
  {239–243} (\bibinfo {year} {2017})}\BibitemShut {NoStop}%
\bibitem [{\citenamefont {Vyaselev}\ \emph {et~al.}(2012)\citenamefont
  {Vyaselev}, \citenamefont {Kato}, \citenamefont {Yamamoto}, \citenamefont
  {Kobayashi}, \citenamefont {Zorina}, \citenamefont {Simonov}, \citenamefont
  {Kushch},\ and\ \citenamefont {Yagubskii}}]{Vyaselev2012}%
  \BibitemOpen
  \bibfield  {author} {\bibinfo {author} {\bibfnamefont {O.~M.}\ \bibnamefont
  {Vyaselev}}, \bibinfo {author} {\bibfnamefont {R.}~\bibnamefont {Kato}},
  \bibinfo {author} {\bibfnamefont {H.~M.}\ \bibnamefont {Yamamoto}}, \bibinfo
  {author} {\bibfnamefont {M.}~\bibnamefont {Kobayashi}}, \bibinfo {author}
  {\bibfnamefont {L.~V.}\ \bibnamefont {Zorina}}, \bibinfo {author}
  {\bibfnamefont {S.~V.}\ \bibnamefont {Simonov}}, \bibinfo {author}
  {\bibfnamefont {N.~D.}\ \bibnamefont {Kushch}}, \ and\ \bibinfo {author}
  {\bibfnamefont {E.~B.}\ \bibnamefont {Yagubskii}},\ }\href {\doibase
  10.3390/cryst2020224} {\bibfield  {journal} {\bibinfo  {journal} {Crystals}\
  }\textbf {\bibinfo {volume} {2}},\ \bibinfo {pages} {224} (\bibinfo {year}
  {2012})}\BibitemShut {NoStop}%
\bibitem [{\citenamefont {Kanoda}(1997)}]{Kanoda1997a}%
  \BibitemOpen
  \bibfield  {author} {\bibinfo {author} {\bibfnamefont {K.}~\bibnamefont
  {Kanoda}},\ }\bibfield  {booktitle} {\emph {\bibinfo {booktitle} {Materials
  and Mechanisms of Superconductivity High Temperature Superconductors V}},\
  }\href
  {http://www.sciencedirect.com/science/article/B6TVJ-3W34K86-5N/2/a92160fcbdfb21cca50446999c5c0648}
  {\bibfield  {journal} {\bibinfo  {journal} {Physica C: Superconductivity}\
  }\textbf {\bibinfo {volume} {282-287}},\ \bibinfo {pages} {299} (\bibinfo
  {year} {1997})}\BibitemShut {NoStop}%
\bibitem [{\citenamefont {Zverev}\ \emph {et~al.}(2019)\citenamefont {Zverev},
  \citenamefont {Biberacher}, \citenamefont {Oberbauer}, \citenamefont
  {Sheikin}, \citenamefont {Alemany}, \citenamefont {Canadell},\ and\
  \citenamefont {Kartsovnik}}]{Zverev2019}%
  \BibitemOpen
  \bibfield  {author} {\bibinfo {author} {\bibfnamefont {V.~N.}\ \bibnamefont
  {Zverev}}, \bibinfo {author} {\bibfnamefont {W.}~\bibnamefont {Biberacher}},
  \bibinfo {author} {\bibfnamefont {S.}~\bibnamefont {Oberbauer}}, \bibinfo
  {author} {\bibfnamefont {I.}~\bibnamefont {Sheikin}}, \bibinfo {author}
  {\bibfnamefont {P.}~\bibnamefont {Alemany}}, \bibinfo {author} {\bibfnamefont
  {E.}~\bibnamefont {Canadell}}, \ and\ \bibinfo {author} {\bibfnamefont
  {M.~V.}\ \bibnamefont {Kartsovnik}},\ }\href {\doibase
  10.1103/PhysRevB.99.125136} {\bibfield  {journal} {\bibinfo  {journal} {Phys.
  Rev. B}\ }\textbf {\bibinfo {volume} {99}},\ \bibinfo {pages} {125136}
  (\bibinfo {year} {2019})}\BibitemShut {NoStop}%
\bibitem [{\citenamefont {Riedl}\ \emph {et~al.}(2022)\citenamefont {Riedl},
  \citenamefont {Gati},\ and\ \citenamefont {Valentí}}]{Riedl2022}%
  \BibitemOpen
  \bibfield  {author} {\bibinfo {author} {\bibfnamefont {K.}~\bibnamefont
  {Riedl}}, \bibinfo {author} {\bibfnamefont {E.}~\bibnamefont {Gati}}, \ and\
  \bibinfo {author} {\bibfnamefont {R.}~\bibnamefont {Valentí}},\ }\href
  {\doibase 10.3390/cryst12121689} {\bibfield  {journal} {\bibinfo  {journal}
  {Crystals}\ }\textbf {\bibinfo {volume} {12}} (\bibinfo {year} {2022}),\
  10.3390/cryst12121689}\BibitemShut {NoStop}%
\bibitem [{\citenamefont {Vyaselev}\ \emph {et~al.}(2011)\citenamefont
  {Vyaselev}, \citenamefont {Kartsovnik}, \citenamefont {Biberacher},
  \citenamefont {Zorina}, \citenamefont {Kushch},\ and\ \citenamefont
  {Yagubskii}}]{Vyaselev2011}%
  \BibitemOpen
  \bibfield  {author} {\bibinfo {author} {\bibfnamefont {O.~M.}\ \bibnamefont
  {Vyaselev}}, \bibinfo {author} {\bibfnamefont {M.~V.}\ \bibnamefont
  {Kartsovnik}}, \bibinfo {author} {\bibfnamefont {W.}~\bibnamefont
  {Biberacher}}, \bibinfo {author} {\bibfnamefont {L.~V.}\ \bibnamefont
  {Zorina}}, \bibinfo {author} {\bibfnamefont {N.~D.}\ \bibnamefont {Kushch}},
  \ and\ \bibinfo {author} {\bibfnamefont {E.~B.}\ \bibnamefont {Yagubskii}},\
  }\href {\doibase 10.1103/PhysRevB.83.094425} {\bibfield  {journal} {\bibinfo
  {journal} {Phys. Rev. B}\ }\textbf {\bibinfo {volume} {83}},\ \bibinfo
  {pages} {094425} (\bibinfo {year} {2011})}\BibitemShut {NoStop}%
\bibitem [{\citenamefont {{O. M. Vyaselev, M. V. Kartsovnik, N. D. Kushch, E.
  B. Yagubskii}}(2012)}]{Vyaselev2012b}%
  \BibitemOpen
  \bibfield  {author} {\bibinfo {author} {\bibnamefont {{O. M. Vyaselev, M. V.
  Kartsovnik, N. D. Kushch, E. B. Yagubskii}}},\ }\href
  {http://dx.doi.org/10.1134/S0021364012110100} {\bibfield  {journal} {\bibinfo
   {journal} {{JETP Letters}}\ }\textbf {\bibinfo {volume} {95}},\ \bibinfo
  {pages} {565} (\bibinfo {year} {2012})}\BibitemShut {NoStop}%
\bibitem [{\citenamefont {M\"uller}\ and\ \citenamefont
  {Thomas}(2018)}]{JMueller2018}%
  \BibitemOpen
  \bibfield  {author} {\bibinfo {author} {\bibfnamefont {J.}~\bibnamefont
  {M\"uller}}\ and\ \bibinfo {author} {\bibfnamefont {T.}~\bibnamefont
  {Thomas}},\ }\href {\doibase 10.3390/cryst8040166} {\bibfield  {journal}
  {\bibinfo  {journal} {Crystals}\ }\textbf {\bibinfo {volume} {8}},\ \bibinfo
  {pages} {166} (\bibinfo {year} {2018})}\BibitemShut {NoStop}%
\bibitem [{\citenamefont {M\"uller}\ \emph {et~al.}(2020)\citenamefont
  {M\"uller}, \citenamefont {Iguchi}, \citenamefont {Taniguchi},\ and\
  \citenamefont {Sasaki}}]{JMueller2020}%
  \BibitemOpen
  \bibfield  {author} {\bibinfo {author} {\bibfnamefont {J.}~\bibnamefont
  {M\"uller}}, \bibinfo {author} {\bibfnamefont {S.}~\bibnamefont {Iguchi}},
  \bibinfo {author} {\bibfnamefont {H.}~\bibnamefont {Taniguchi}}, \ and\
  \bibinfo {author} {\bibfnamefont {T.}~\bibnamefont {Sasaki}},\ }\href
  {\doibase 10.1103/PhysRevB.102.100103} {\bibfield  {journal} {\bibinfo
  {journal} {Phys. Rev. B}\ }\textbf {\bibinfo {volume} {102}},\ \bibinfo
  {pages} {100103} (\bibinfo {year} {2020})}\BibitemShut {NoStop}%
\bibitem [{\citenamefont {Raquet}\ \emph {et~al.}(2000)\citenamefont {Raquet},
  \citenamefont {Anane}, \citenamefont {Wirth}, \citenamefont {Xiong},\ and\
  \citenamefont {von Moln\'ar}}]{Raquet2000}%
  \BibitemOpen
  \bibfield  {author} {\bibinfo {author} {\bibfnamefont {B.}~\bibnamefont
  {Raquet}}, \bibinfo {author} {\bibfnamefont {A.}~\bibnamefont {Anane}},
  \bibinfo {author} {\bibfnamefont {S.}~\bibnamefont {Wirth}}, \bibinfo
  {author} {\bibfnamefont {P.}~\bibnamefont {Xiong}}, \ and\ \bibinfo {author}
  {\bibfnamefont {S.}~\bibnamefont {von Moln\'ar}},\ }\href {\doibase
  10.1103/PhysRevLett.84.4485} {\bibfield  {journal} {\bibinfo  {journal}
  {Physical Review Letters}\ }\textbf {\bibinfo {volume} {84}},\ \bibinfo
  {pages} {4485} (\bibinfo {year} {2000})}\BibitemShut {NoStop}%
\bibitem [{\citenamefont {Lang}\ \emph {et~al.}(2014)\citenamefont {Lang},
  \citenamefont {Lunkenheimer}, \citenamefont {M\"uller}, \citenamefont
  {Loidl}, \citenamefont {Hartmann}, \citenamefont {Hoang}, \citenamefont
  {Gati}, \citenamefont {Schubert},\ and\ \citenamefont
  {Schlueter}}]{Lang2014}%
  \BibitemOpen
  \bibfield  {author} {\bibinfo {author} {\bibfnamefont {M.}~\bibnamefont
  {Lang}}, \bibinfo {author} {\bibfnamefont {P.}~\bibnamefont {Lunkenheimer}},
  \bibinfo {author} {\bibfnamefont {J.}~\bibnamefont {M\"uller}}, \bibinfo
  {author} {\bibfnamefont {A.}~\bibnamefont {Loidl}}, \bibinfo {author}
  {\bibfnamefont {B.}~\bibnamefont {Hartmann}}, \bibinfo {author}
  {\bibfnamefont {N.~H.}\ \bibnamefont {Hoang}}, \bibinfo {author}
  {\bibfnamefont {E.}~\bibnamefont {Gati}}, \bibinfo {author} {\bibfnamefont
  {H.}~\bibnamefont {Schubert}}, \ and\ \bibinfo {author} {\bibfnamefont
  {J.~A.}\ \bibnamefont {Schlueter}},\ }\href {\doibase
  10.1109/TMAG.2013.2296333} {\bibfield  {journal} {\bibinfo  {journal} {IEEE
  Transactions on Magnetics}\ }\textbf {\bibinfo {volume} {50}},\ \bibinfo
  {pages} {1} (\bibinfo {year} {2014})}\BibitemShut {NoStop}%
\bibitem [{\citenamefont {Lunkenheimer}\ \emph {et~al.}(2009)\citenamefont
  {Lunkenheimer}, \citenamefont {Krohns}, \citenamefont {Riegg}, \citenamefont
  {Ebbinghaus}, \citenamefont {Reller},\ and\ \citenamefont
  {Loidl}}]{Lunkenheimer2009}%
  \BibitemOpen
  \bibfield  {author} {\bibinfo {author} {\bibfnamefont {P.}~\bibnamefont
  {Lunkenheimer}}, \bibinfo {author} {\bibfnamefont {S.}~\bibnamefont
  {Krohns}}, \bibinfo {author} {\bibfnamefont {S.}~\bibnamefont {Riegg}},
  \bibinfo {author} {\bibfnamefont {S.}~\bibnamefont {Ebbinghaus}}, \bibinfo
  {author} {\bibfnamefont {A.}~\bibnamefont {Reller}}, \ and\ \bibinfo {author}
  {\bibfnamefont {A.}~\bibnamefont {Loidl}},\ }\href {\doibase
  10.1140/epjst/e2010-01212-5} {\bibfield  {journal} {\bibinfo  {journal} {The
  European Physical Journal Special Topics}\ }\textbf {\bibinfo {volume}
  {180}},\ \bibinfo {pages} {61} (\bibinfo {year} {2009})}\BibitemShut
  {NoStop}%
\bibitem [{\citenamefont {Bobnar}\ \emph {et~al.}(2002)\citenamefont {Bobnar},
  \citenamefont {Lunkenheimer}, \citenamefont {Paraskevopoulos},\ and\
  \citenamefont {Loidl}}]{Bobnar2002}%
  \BibitemOpen
  \bibfield  {author} {\bibinfo {author} {\bibfnamefont {V.}~\bibnamefont
  {Bobnar}}, \bibinfo {author} {\bibfnamefont {P.}~\bibnamefont
  {Lunkenheimer}}, \bibinfo {author} {\bibfnamefont {M.}~\bibnamefont
  {Paraskevopoulos}}, \ and\ \bibinfo {author} {\bibfnamefont {A.}~\bibnamefont
  {Loidl}},\ }\href@noop {} {\bibfield  {journal} {\bibinfo  {journal}
  {Physical Review B}\ }\textbf {\bibinfo {volume} {65}},\ \bibinfo {pages}
  {184403} (\bibinfo {year} {2002})}\BibitemShut {NoStop}%
\bibitem [{\citenamefont {Cross}(1987)}]{Cross1987}%
  \BibitemOpen
  \bibfield  {author} {\bibinfo {author} {\bibfnamefont {L.~E.}\ \bibnamefont
  {Cross}},\ }\href {\doibase 10.1080/00150198708016945} {\bibfield  {journal}
  {\bibinfo  {journal} {Ferroelectrics}\ }\textbf {\bibinfo {volume} {76}},\
  \bibinfo {pages} {241} (\bibinfo {year} {1987})}\BibitemShut {NoStop}%
\bibitem [{\citenamefont {Samara}(2003)}]{Samara2003}%
  \BibitemOpen
  \bibfield  {author} {\bibinfo {author} {\bibfnamefont {G.~A.}\ \bibnamefont
  {Samara}},\ }\href {\doibase 10.1088/0953-8984/15/9/202} {\bibfield
  {journal} {\bibinfo  {journal} {Journal of Physics: Condensed Matter}\
  }\textbf {\bibinfo {volume} {15}},\ \bibinfo {pages} {R367} (\bibinfo {year}
  {2003})}\BibitemShut {NoStop}%
\bibitem [{\citenamefont {Viehland}\ \emph {et~al.}(1990)\citenamefont
  {Viehland}, \citenamefont {Jang}, \citenamefont {Cross},\ and\ \citenamefont
  {Wuttig}}]{Viehland1990}%
  \BibitemOpen
  \bibfield  {author} {\bibinfo {author} {\bibfnamefont {D.}~\bibnamefont
  {Viehland}}, \bibinfo {author} {\bibfnamefont {S.~J.}\ \bibnamefont {Jang}},
  \bibinfo {author} {\bibfnamefont {L.~E.}\ \bibnamefont {Cross}}, \ and\
  \bibinfo {author} {\bibfnamefont {M.}~\bibnamefont {Wuttig}},\ }\href
  {\doibase 10.1063/1.346425} {\bibfield  {journal} {\bibinfo  {journal} {J.
  Appl. Phys.}\ }\textbf {\bibinfo {volume} {68}},\ \bibinfo {pages} {2916}
  (\bibinfo {year} {1990})}\BibitemShut {NoStop}%
\bibitem [{\citenamefont {Fischer}\ \emph {et~al.}(2021)\citenamefont
  {Fischer}, \citenamefont {D'Avino}, \citenamefont {Masino}, \citenamefont
  {Mezzadri}, \citenamefont {Lunkenheimer}, \citenamefont {Soos},\ and\
  \citenamefont {Girlando}}]{Fischer2021}%
  \BibitemOpen
  \bibfield  {author} {\bibinfo {author} {\bibfnamefont {J.~K.~H.}\
  \bibnamefont {Fischer}}, \bibinfo {author} {\bibfnamefont {G.}~\bibnamefont
  {D'Avino}}, \bibinfo {author} {\bibfnamefont {M.}~\bibnamefont {Masino}},
  \bibinfo {author} {\bibfnamefont {F.}~\bibnamefont {Mezzadri}}, \bibinfo
  {author} {\bibfnamefont {P.}~\bibnamefont {Lunkenheimer}}, \bibinfo {author}
  {\bibfnamefont {Z.~G.}\ \bibnamefont {Soos}}, \ and\ \bibinfo {author}
  {\bibfnamefont {A.}~\bibnamefont {Girlando}},\ }\href@noop {} {\bibfield
  {journal} {\bibinfo  {journal} {Physical Review B}\ }\textbf {\bibinfo
  {volume} {103}},\ \bibinfo {pages} {115104} (\bibinfo {year}
  {2021})}\BibitemShut {NoStop}%
\bibitem [{\citenamefont {Canossa}\ \emph {et~al.}(2021)\citenamefont
  {Canossa}, \citenamefont {Ferrari}, \citenamefont {Sippel}, \citenamefont
  {Fischer}, \citenamefont {Pfattner}, \citenamefont {Frison}, \citenamefont
  {Masino}, \citenamefont {Mas-Torrent}, \citenamefont {Lunkenheimer},
  \citenamefont {Rovira},\ and\ \citenamefont {Girlando}}]{Canossa2021}%
  \BibitemOpen
  \bibfield  {author} {\bibinfo {author} {\bibfnamefont {S.}~\bibnamefont
  {Canossa}}, \bibinfo {author} {\bibfnamefont {E.}~\bibnamefont {Ferrari}},
  \bibinfo {author} {\bibfnamefont {P.}~\bibnamefont {Sippel}}, \bibinfo
  {author} {\bibfnamefont {J.~K.~H.}\ \bibnamefont {Fischer}}, \bibinfo
  {author} {\bibfnamefont {R.}~\bibnamefont {Pfattner}}, \bibinfo {author}
  {\bibfnamefont {R.}~\bibnamefont {Frison}}, \bibinfo {author} {\bibfnamefont
  {M.}~\bibnamefont {Masino}}, \bibinfo {author} {\bibfnamefont
  {M.}~\bibnamefont {Mas-Torrent}}, \bibinfo {author} {\bibfnamefont
  {P.}~\bibnamefont {Lunkenheimer}}, \bibinfo {author} {\bibfnamefont
  {C.}~\bibnamefont {Rovira}}, \ and\ \bibinfo {author} {\bibfnamefont
  {A.}~\bibnamefont {Girlando}},\ }\href@noop {} {\bibfield  {journal}
  {\bibinfo  {journal} {The Journal of Physical Chemistry C}\ }\textbf
  {\bibinfo {volume} {125}},\ \bibinfo {pages} {25816} (\bibinfo {year}
  {2021})}\BibitemShut {NoStop}%
\bibitem [{\citenamefont {Thurn}\ \emph {et~al.}(2021)\citenamefont {Thurn},
  \citenamefont {Eibisch}, \citenamefont {Ata}, \citenamefont {Winkler},
  \citenamefont {Lunkenheimer}, \citenamefont {K{\'e}zsm{\'a}rki},
  \citenamefont {Tutsch}, \citenamefont {Saito}, \citenamefont {Hartmann},
  \citenamefont {Zimmermann}, \citenamefont {Hanna}, \citenamefont {Islam},
  \citenamefont {Chillal}, \citenamefont {Lake}, \citenamefont {Wolf},\ and\
  \citenamefont {Lang}}]{Thurn2021}%
  \BibitemOpen
  \bibfield  {author} {\bibinfo {author} {\bibfnamefont {C.}~\bibnamefont
  {Thurn}}, \bibinfo {author} {\bibfnamefont {P.}~\bibnamefont {Eibisch}},
  \bibinfo {author} {\bibfnamefont {A.}~\bibnamefont {Ata}}, \bibinfo {author}
  {\bibfnamefont {M.}~\bibnamefont {Winkler}}, \bibinfo {author} {\bibfnamefont
  {P.}~\bibnamefont {Lunkenheimer}}, \bibinfo {author} {\bibfnamefont
  {I.}~\bibnamefont {K{\'e}zsm{\'a}rki}}, \bibinfo {author} {\bibfnamefont
  {U.}~\bibnamefont {Tutsch}}, \bibinfo {author} {\bibfnamefont
  {Y.}~\bibnamefont {Saito}}, \bibinfo {author} {\bibfnamefont
  {S.}~\bibnamefont {Hartmann}}, \bibinfo {author} {\bibfnamefont
  {J.}~\bibnamefont {Zimmermann}}, \bibinfo {author} {\bibfnamefont {A.~R.~N.}\
  \bibnamefont {Hanna}}, \bibinfo {author} {\bibfnamefont {A.~T. M.~N.}\
  \bibnamefont {Islam}}, \bibinfo {author} {\bibfnamefont {S.}~\bibnamefont
  {Chillal}}, \bibinfo {author} {\bibfnamefont {B.}~\bibnamefont {Lake}},
  \bibinfo {author} {\bibfnamefont {B.}~\bibnamefont {Wolf}}, \ and\ \bibinfo
  {author} {\bibfnamefont {M.}~\bibnamefont {Lang}},\ }\href {\doibase
  10.1038/s41535-021-00395-6} {\bibfield  {journal} {\bibinfo  {journal} {npj
  Quantum Materials}\ }\textbf {\bibinfo {volume} {6}},\ \bibinfo {pages} {95}
  (\bibinfo {year} {2021})}\BibitemShut {NoStop}%
\bibitem [{\citenamefont {Ward}\ \emph {et~al.}(2009)\citenamefont {Ward},
  \citenamefont {Zhang}, \citenamefont {Yin}, \citenamefont {Zhang},
  \citenamefont {Liu}, \citenamefont {Snijders}, \citenamefont {Jesse},
  \citenamefont {Plummer}, \citenamefont {Cheng}, \citenamefont {Dagotto},\
  and\ \citenamefont {Shen}}]{Ward2009}%
  \BibitemOpen
  \bibfield  {author} {\bibinfo {author} {\bibfnamefont {T.~Z.}\ \bibnamefont
  {Ward}}, \bibinfo {author} {\bibfnamefont {X.~G.}\ \bibnamefont {Zhang}},
  \bibinfo {author} {\bibfnamefont {L.~F.}\ \bibnamefont {Yin}}, \bibinfo
  {author} {\bibfnamefont {X.~Q.}\ \bibnamefont {Zhang}}, \bibinfo {author}
  {\bibfnamefont {M.}~\bibnamefont {Liu}}, \bibinfo {author} {\bibfnamefont
  {P.~C.}\ \bibnamefont {Snijders}}, \bibinfo {author} {\bibfnamefont
  {S.}~\bibnamefont {Jesse}}, \bibinfo {author} {\bibfnamefont {E.~W.}\
  \bibnamefont {Plummer}}, \bibinfo {author} {\bibfnamefont {Z.~H.}\
  \bibnamefont {Cheng}}, \bibinfo {author} {\bibfnamefont {E.}~\bibnamefont
  {Dagotto}}, \ and\ \bibinfo {author} {\bibfnamefont {J.}~\bibnamefont
  {Shen}},\ }\href {\doibase 10.1103/PhysRevLett.102.087201} {\bibfield
  {journal} {\bibinfo  {journal} {Phys. Rev. Lett.}\ }\textbf {\bibinfo
  {volume} {102}},\ \bibinfo {pages} {087201} (\bibinfo {year}
  {2009})}\BibitemShut {NoStop}%
\bibitem [{\citenamefont {{Yu}}(2004)}]{Yu2004b}%
  \BibitemOpen
  \bibfield  {author} {\bibinfo {author} {\bibfnamefont {C.~C.}\ \bibnamefont
  {{Yu}}},\ }\href@noop {} {\bibfield  {journal} {\bibinfo  {journal} {Journal
  of Low Temperature Physics}\ }\textbf {\bibinfo {volume} {137}},\ \bibinfo
  {pages} {251} (\bibinfo {year} {2004})}\BibitemShut {NoStop}%
\bibitem [{\citenamefont {Weissman}(1988)}]{Weissman1988}%
  \BibitemOpen
  \bibfield  {author} {\bibinfo {author} {\bibfnamefont {M.~B.}\ \bibnamefont
  {Weissman}},\ }\href {\doibase 10.1103/RevModPhys.60.537} {\bibfield
  {journal} {\bibinfo  {journal} {Reviews of Modern Physics}\ }\textbf
  {\bibinfo {volume} {60}},\ \bibinfo {pages} {537} (\bibinfo {year}
  {1988})}\BibitemShut {NoStop}%
\bibitem [{\citenamefont {Chen}\ and\ \citenamefont {Yu}(2007)}]{Chen2007}%
  \BibitemOpen
  \bibfield  {author} {\bibinfo {author} {\bibfnamefont {Z.}~\bibnamefont
  {Chen}}\ and\ \bibinfo {author} {\bibfnamefont {C.}~\bibnamefont {Yu}},\
  }\href {\doibase 10.1103/PhysRevLett.98.057204} {\bibfield  {journal}
  {\bibinfo  {journal} {Phys. Rev. Lett.}\ }\textbf {\bibinfo {volume} {98}},\
  \bibinfo {pages} {057204} (\bibinfo {year} {2007})}\BibitemShut {NoStop}%
\bibitem [{\citenamefont {Daptary}\ \emph {et~al.}(2019)\citenamefont
  {Daptary}, \citenamefont {Kumar}, \citenamefont {Kareev}, \citenamefont
  {Chakhalian}, \citenamefont {Bid},\ and\ \citenamefont
  {Middey}}]{Daptary2019}%
  \BibitemOpen
  \bibfield  {author} {\bibinfo {author} {\bibfnamefont {G.~N.}\ \bibnamefont
  {Daptary}}, \bibinfo {author} {\bibfnamefont {S.}~\bibnamefont {Kumar}},
  \bibinfo {author} {\bibfnamefont {M.}~\bibnamefont {Kareev}}, \bibinfo
  {author} {\bibfnamefont {J.}~\bibnamefont {Chakhalian}}, \bibinfo {author}
  {\bibfnamefont {A.}~\bibnamefont {Bid}}, \ and\ \bibinfo {author}
  {\bibfnamefont {S.}~\bibnamefont {Middey}},\ }\href {\doibase
  10.1103/PhysRevB.100.125105} {\bibfield  {journal} {\bibinfo  {journal}
  {Phys. Rev. B}\ }\textbf {\bibinfo {volume} {100}},\ \bibinfo {pages}
  {125105} (\bibinfo {year} {2019})}\BibitemShut {NoStop}%
\bibitem [{\citenamefont {Pinteri\v{c}}\ \emph {et~al.}(2014)\citenamefont
  {Pinteri\v{c}}, \citenamefont {\v{C}ulo}, \citenamefont {Milat},
  \citenamefont {Basleti\v{c}}, \citenamefont {Korin-Hamzi\v{c}}, \citenamefont
  {Tafra}, \citenamefont {Hamzi\v{c}}, \citenamefont {Ivek}, \citenamefont
  {Peterseim},\ and\ \citenamefont {Miyagawa}}]{Pinteric2014}%
  \BibitemOpen
  \bibfield  {author} {\bibinfo {author} {\bibfnamefont {M.}~\bibnamefont
  {Pinteri\v{c}}}, \bibinfo {author} {\bibfnamefont {M.}~\bibnamefont
  {\v{C}ulo}}, \bibinfo {author} {\bibfnamefont {O.}~\bibnamefont {Milat}},
  \bibinfo {author} {\bibfnamefont {M.}~\bibnamefont {Basleti\v{c}}}, \bibinfo
  {author} {\bibfnamefont {B.}~\bibnamefont {Korin-Hamzi\v{c}}}, \bibinfo
  {author} {\bibfnamefont {E.}~\bibnamefont {Tafra}}, \bibinfo {author}
  {\bibfnamefont {A.}~\bibnamefont {Hamzi\v{c}}}, \bibinfo {author}
  {\bibfnamefont {T.}~\bibnamefont {Ivek}}, \bibinfo {author} {\bibfnamefont
  {T.}~\bibnamefont {Peterseim}}, \ and\ \bibinfo {author} {\bibfnamefont
  {K.}~\bibnamefont {Miyagawa}},\ }\href@noop {} {\bibfield  {journal}
  {\bibinfo  {journal} {Phys. Rev. B}\ }\textbf {\bibinfo {volume} {90}},\
  \bibinfo {pages} {195139} (\bibinfo {year} {2014})}\BibitemShut {NoStop}%
\bibitem [{\citenamefont {Fisher}\ and\ \citenamefont
  {Huse}(1988{\natexlab{a}})}]{Fisher1988a}%
  \BibitemOpen
  \bibfield  {author} {\bibinfo {author} {\bibfnamefont {D.~S.}\ \bibnamefont
  {Fisher}}\ and\ \bibinfo {author} {\bibfnamefont {D.~A.}\ \bibnamefont
  {Huse}},\ }\href {\doibase 10.1103/PhysRevB.38.373} {\bibfield  {journal}
  {\bibinfo  {journal} {Phys. Rev. B}\ }\textbf {\bibinfo {volume} {38}},\
  \bibinfo {pages} {373} (\bibinfo {year} {1988}{\natexlab{a}})}\BibitemShut
  {NoStop}%
\bibitem [{\citenamefont {Fisher}\ and\ \citenamefont
  {Huse}(1988{\natexlab{b}})}]{Fisher1988}%
  \BibitemOpen
  \bibfield  {author} {\bibinfo {author} {\bibfnamefont {D.~S.}\ \bibnamefont
  {Fisher}}\ and\ \bibinfo {author} {\bibfnamefont {D.~A.}\ \bibnamefont
  {Huse}},\ }\href@noop {} {\bibfield  {journal} {\bibinfo  {journal} {Phys.
  Rev. B}\ }\textbf {\bibinfo {volume} {38}},\ \bibinfo {pages} {386} (\bibinfo
  {year} {1988}{\natexlab{b}})}\BibitemShut {NoStop}%
\bibitem [{\citenamefont {Weissman}\ \emph {et~al.}(1992)\citenamefont
  {Weissman}, \citenamefont {Israeloff},\ and\ \citenamefont
  {Alers}}]{Weissman1992}%
  \BibitemOpen
  \bibfield  {author} {\bibinfo {author} {\bibfnamefont {M.~B.}\ \bibnamefont
  {Weissman}}, \bibinfo {author} {\bibfnamefont {N.~E.}\ \bibnamefont
  {Israeloff}}, \ and\ \bibinfo {author} {\bibfnamefont {G.~B.}\ \bibnamefont
  {Alers}},\ }\href
  {http://www.sciencedirect.com/science/article/B6TJJ-46G6212-147/2/23e8f0d49caf0808959ec8b0238d719a}
  {\bibfield  {journal} {\bibinfo  {journal} {Journal of Magnetism and Magnetic
  Materials}\ }\textbf {\bibinfo {volume} {114}},\ \bibinfo {pages} {87}
  (\bibinfo {year} {1992})}\BibitemShut {NoStop}%
\bibitem [{\citenamefont {Weissman}(1993)}]{Weissman1993}%
  \BibitemOpen
  \bibfield  {author} {\bibinfo {author} {\bibfnamefont {M.~B.}\ \bibnamefont
  {Weissman}},\ }\href {\doibase 10.1103/RevModPhys.65.829} {\bibfield
  {journal} {\bibinfo  {journal} {Reviews of Modern Physics}\ }\textbf
  {\bibinfo {volume} {65}},\ \bibinfo {pages} {829} (\bibinfo {year}
  {1993})}\BibitemShut {NoStop}%
\bibitem [{\citenamefont {Russell}\ and\ \citenamefont
  {Israeloff}(2000)}]{Russell2000}%
  \BibitemOpen
  \bibfield  {author} {\bibinfo {author} {\bibfnamefont {E.~V.}\ \bibnamefont
  {Russell}}\ and\ \bibinfo {author} {\bibfnamefont {N.~E.}\ \bibnamefont
  {Israeloff}},\ }\href {\doibase 10.1038/35047037} {\bibfield  {journal}
  {\bibinfo  {journal} {Nature}\ }\textbf {\bibinfo {volume} {408}},\ \bibinfo
  {pages} {695} (\bibinfo {year} {2000})}\BibitemShut {NoStop}%
\bibitem [{\citenamefont {M\"{u}ller}(2011)}]{JMueller2011}%
  \BibitemOpen
  \bibfield  {author} {\bibinfo {author} {\bibfnamefont {J.}~\bibnamefont
  {M\"{u}ller}},\ }\href {\doibase 10.1002/cphc.201000814} {\bibfield
  {journal} {\bibinfo  {journal} {ChemPhysChem}\ }\textbf {\bibinfo {volume}
  {12}},\ \bibinfo {pages} {1222} (\bibinfo {year} {2011})}\BibitemShut
  {NoStop}%
\bibitem [{\citenamefont {Pott}\ and\ \citenamefont
  {Schefzyk}(1983)}]{Pott1983}%
  \BibitemOpen
  \bibfield  {author} {\bibinfo {author} {\bibfnamefont {R.}~\bibnamefont
  {Pott}}\ and\ \bibinfo {author} {\bibfnamefont {R.}~\bibnamefont
  {Schefzyk}},\ }\href {\doibase 10.1088/0022-3735/16/5/018} {\bibfield
  {journal} {\bibinfo  {journal} {Journal of Physics E: Scientific
  Instruments}\ }\textbf {\bibinfo {volume} {16}},\ \bibinfo {pages} {444}
  (\bibinfo {year} {1983})}\BibitemShut {NoStop}%
\bibitem [{\citenamefont {B\"ohmer}\ \emph {et~al.}(1989)\citenamefont
  {B\"ohmer}, \citenamefont {Maglione}, \citenamefont {Lunkenheimer},\ and\
  \citenamefont {Loidl}}]{Boehmer1989}%
  \BibitemOpen
  \bibfield  {author} {\bibinfo {author} {\bibfnamefont {R.}~\bibnamefont
  {B\"ohmer}}, \bibinfo {author} {\bibfnamefont {M.}~\bibnamefont {Maglione}},
  \bibinfo {author} {\bibfnamefont {P.}~\bibnamefont {Lunkenheimer}}, \ and\
  \bibinfo {author} {\bibfnamefont {A.}~\bibnamefont {Loidl}},\ }\href@noop {}
  {\bibfield  {journal} {\bibinfo  {journal} {Journal of Applied Physics}\
  }\textbf {\bibinfo {volume} {65}},\ \bibinfo {pages} {901} (\bibinfo {year}
  {1989})}\BibitemShut {NoStop}%
\bibitem [{\citenamefont {Macutkevic}\ \emph {et~al.}(2011)\citenamefont
  {Macutkevic}, \citenamefont {Banys}, \citenamefont {Bussmann-Holder},\ and\
  \citenamefont {Bishop}}]{Macutkevic2011}%
  \BibitemOpen
  \bibfield  {author} {\bibinfo {author} {\bibfnamefont {J.}~\bibnamefont
  {Macutkevic}}, \bibinfo {author} {\bibfnamefont {J.}~\bibnamefont {Banys}},
  \bibinfo {author} {\bibfnamefont {A.}~\bibnamefont {Bussmann-Holder}}, \ and\
  \bibinfo {author} {\bibfnamefont {A.~R.}\ \bibnamefont {Bishop}},\ }\href
  {\doibase 10.1103/PhysRevB.83.184301} {\bibfield  {journal} {\bibinfo
  {journal} {Phys. Rev. B}\ }\textbf {\bibinfo {volume} {83}},\ \bibinfo
  {pages} {184301} (\bibinfo {year} {2011})}\BibitemShut {NoStop}%
\bibitem [{\citenamefont {Lunkenheimer}\ \emph {et~al.}(1995)\citenamefont
  {Lunkenheimer}, \citenamefont {Knebel}, \citenamefont {Pimenov},
  \citenamefont {Emel'chenko},\ and\ \citenamefont {Loidl}}]{Lunkenheimer1995}%
  \BibitemOpen
  \bibfield  {author} {\bibinfo {author} {\bibfnamefont {P.}~\bibnamefont
  {Lunkenheimer}}, \bibinfo {author} {\bibfnamefont {G.}~\bibnamefont
  {Knebel}}, \bibinfo {author} {\bibfnamefont {A.}~\bibnamefont {Pimenov}},
  \bibinfo {author} {\bibfnamefont {G.~A.}\ \bibnamefont {Emel'chenko}}, \ and\
  \bibinfo {author} {\bibfnamefont {A.}~\bibnamefont {Loidl}},\ }\href@noop {}
  {\bibfield  {journal} {\bibinfo  {journal} {Zeitschrift f\"ur Physik B
  Condensed Matter}\ }\textbf {\bibinfo {volume} {99}},\ \bibinfo {pages} {507}
  (\bibinfo {year} {1995})}\BibitemShut {NoStop}%
\bibitem [{\citenamefont {Lunkenheimer}\ \emph {et~al.}(2002)\citenamefont
  {Lunkenheimer}, \citenamefont {Bobnar}, \citenamefont {Pronin}, \citenamefont
  {Ritus}, \citenamefont {Volkov},\ and\ \citenamefont
  {Loidl}}]{Lunkenheimer2002}%
  \BibitemOpen
  \bibfield  {author} {\bibinfo {author} {\bibfnamefont {P.}~\bibnamefont
  {Lunkenheimer}}, \bibinfo {author} {\bibfnamefont {V.}~\bibnamefont
  {Bobnar}}, \bibinfo {author} {\bibfnamefont {A.~V.}\ \bibnamefont {Pronin}},
  \bibinfo {author} {\bibfnamefont {A.~I.}\ \bibnamefont {Ritus}}, \bibinfo
  {author} {\bibfnamefont {A.~A.}\ \bibnamefont {Volkov}}, \ and\ \bibinfo
  {author} {\bibfnamefont {A.}~\bibnamefont {Loidl}},\ }\href@noop {}
  {\bibfield  {journal} {\bibinfo  {journal} {Physical Review B}\ }\textbf
  {\bibinfo {volume} {66}},\ \bibinfo {pages} {052105} (\bibinfo {year}
  {2002})}\BibitemShut {NoStop}%
\bibitem [{\citenamefont {Lunkenheimer}\ \emph {et~al.}(1996)\citenamefont
  {Lunkenheimer}, \citenamefont {Pimenov}, \citenamefont {Schiener},
  \citenamefont {B\"ohmer},\ and\ \citenamefont {Loidl}}]{Lunkenheimer1996}%
  \BibitemOpen
  \bibfield  {author} {\bibinfo {author} {\bibfnamefont {P.}~\bibnamefont
  {Lunkenheimer}}, \bibinfo {author} {\bibfnamefont {A.}~\bibnamefont
  {Pimenov}}, \bibinfo {author} {\bibfnamefont {B.}~\bibnamefont {Schiener}},
  \bibinfo {author} {\bibfnamefont {R.}~\bibnamefont {B\"ohmer}}, \ and\
  \bibinfo {author} {\bibfnamefont {A.}~\bibnamefont {Loidl}},\ }\href
  {\doibase 10.1209/epl/i1996-00387-4} {\bibfield  {journal} {\bibinfo
  {journal} {Europhysics Letters}\ }\textbf {\bibinfo {volume} {33}},\ \bibinfo
  {pages} {611} (\bibinfo {year} {1996})}\BibitemShut {NoStop}%
\bibitem [{\citenamefont {Emmert}\ \emph {et~al.}(2011)\citenamefont {Emmert},
  \citenamefont {Wolf}, \citenamefont {Gulich}, \citenamefont {Krohns},
  \citenamefont {Kastner}, \citenamefont {Lunkenheimer},\ and\ \citenamefont
  {Loidl}}]{Emmert2011}%
  \BibitemOpen
  \bibfield  {author} {\bibinfo {author} {\bibfnamefont {S.}~\bibnamefont
  {Emmert}}, \bibinfo {author} {\bibfnamefont {M.}~\bibnamefont {Wolf}},
  \bibinfo {author} {\bibfnamefont {R.}~\bibnamefont {Gulich}}, \bibinfo
  {author} {\bibfnamefont {S.}~\bibnamefont {Krohns}}, \bibinfo {author}
  {\bibfnamefont {S.}~\bibnamefont {Kastner}}, \bibinfo {author} {\bibfnamefont
  {P.}~\bibnamefont {Lunkenheimer}}, \ and\ \bibinfo {author} {\bibfnamefont
  {A.}~\bibnamefont {Loidl}},\ }\href {\doibase 10.1140/epjb/e2011-20439-8}
  {\bibfield  {journal} {\bibinfo  {journal} {The European Physical Journal B}\
  }\textbf {\bibinfo {volume} {83}},\ \bibinfo {pages} {157} (\bibinfo {year}
  {2011})}\BibitemShut {NoStop}%
\bibitem [{\citenamefont {Cole}\ and\ \citenamefont {Cole}(1941)}]{Cole1941}%
  \BibitemOpen
  \bibfield  {author} {\bibinfo {author} {\bibfnamefont {K.~S.}\ \bibnamefont
  {Cole}}\ and\ \bibinfo {author} {\bibfnamefont {R.~H.}\ \bibnamefont
  {Cole}},\ }\href {\doibase 10.1063/1.1750906} {\bibfield  {journal} {\bibinfo
   {journal} {The Journal of Chemical Physics}\ }\textbf {\bibinfo {volume}
  {9}},\ \bibinfo {pages} {341} (\bibinfo {year} {1941})}\BibitemShut {NoStop}%
\bibitem [{\citenamefont {Mott}\ and\ \citenamefont {Davis}(1979)}]{Mott1979}%
  \BibitemOpen
  \bibfield  {author} {\bibinfo {author} {\bibfnamefont {N.~F.}\ \bibnamefont
  {Mott}}\ and\ \bibinfo {author} {\bibfnamefont {E.~A.}\ \bibnamefont
  {Davis}},\ }\href@noop {} {\emph {\bibinfo {title} {{Electronic Processes in
  Non-Crystalline Materials}}}}\ (\bibinfo  {publisher} {Clarendon Press,
  Oxford},\ \bibinfo {year} {1979})\BibitemShut {NoStop}%
\end{thebibliography}%

\end{document}